\documentstyle[aas2pp4]{article}
\newcommand{\kms}{km ${\rm s}^{-1}$\,}
\newcommand{\vsini}{$v$\,sin\,$i$}

\newcommand{\ergs}{erg~s$^{-1}$}

\newcommand{\lx}{$L_{X}$\,}

\newcommand{\lbol}{$L_{\rm bol}$}
\newcommand{\lsol}{$L_{\odot}$}
\newcommand{\msol}{$M_{\odot}$}
\newcommand{\vesc}{$v_{esc}$}

\newcommand{\bj}{$B_J$\,}
\newcommand{\jminusf}{($B_{J}-F$)\,}
\newcommand{\BV}{($B-V$)}

\newcommand{\dudvdw}{$\sigma_U$, $\sigma_V$, $\sigma_W$}
\newcommand{\dxdydz}{$\sigma_X$, $\sigma_Y$, $\sigma_Z$}

\newcommand{\rosat}{{\it ROSAT}\/}
\newcommand{\iras}{{\it IRAS}\/}
\newcommand{\prems}{{pre-MS}\/}

\received{December 9, 1999}
\lefthead{Mamajek et al.}
\righthead{$\eta$ Cha Cluster}

\begin{document}

\title{The $\eta$ Chamaeleontis Cluster: Origin in the Sco-Cen OB Association}

\author{Eric E. Mamajek\altaffilmark{1,2,3},
        Warrick A. Lawson\altaffilmark{1} and 
        Eric D. Feigelson\altaffilmark{4}}
\altaffiltext{1}{School of Physics, University College, University of
New South Wales, Australian Defence  Force Academy, Canberra ACT 2600, 
Australia}
\altaffiltext{2}{J. William Fulbright Fellow}
\altaffiltext{3}{Current address: Steward Observatory, 
Department of Astronomy, University of Arizona, 933 N. Cherry Ave., 
Tucson AZ 85721, USA}
\altaffiltext{4}{Department of Astronomy \& Astrophysics, Pennsylvania
State University, University Park, PA 16802, USA}

\received{1999 Dec 9}

\begin{abstract}
\normalsize
A young, nearby compact aggregate of X-ray emitting pre-main sequence
stars was recently discovered in the vicinity of $\eta$ Chamaeleontis
(Mamajek, Lawson \& Feigelson 1999, ApJ, 516, L77).  In this paper, we
further investigate this cluster: its membership, its environs
and origins. \rosat\/ High-Resolution Imager X-ray data for the
cluster's T Tauri stars show high levels of magnetic activity and 
variability. The cluster has an anomalous X-ray luminosity function
compared to other young clusters, deficient in stars with low, 
but detectable X-ray luminosities. This suggests that many low-mass 
members have escaped the surveyed core region. Photographic photometry from
the USNO-A2.0 catalog indicates that additional, X-ray-quiet members exist 
in the cluster core region. 
The components of the eclipsing binary RS Cha, previously
modeled in the literature as post-main sequence with discordant ages,
are shown to be consistent with being coeval \prems\, stars.

We compute the Galactic motion of the cluster from {\it
Hipparcos\,} data, and compare it to other young stars and associations
in the fourth Galactic quadrant.  The kinematic study shows that the
$\eta$ Cha cluster, as well as members of the TW Hya association and
a new group near $\epsilon$ Cha, probably originated near the giant
molecular cloud complex that formed the two oldest subgroups of the
Sco-Cen OB association roughly 10-15 Myr ago.  
Their dispersal is consistent with the velocity dispersions seen
in giant molecular clouds.  A large H {\small I} filament and dust lane
located near $\eta$ Cha has been identified as part of a superbubble
formed by Sco-Cen OB winds and supernova remnants.  The passage of the
superbubble may have terminated star-formation in the $\eta$ Cha
cluster and dispersed its natal molecular gas. \\

\end{abstract}


\section{Introduction \label{intro}}

Intermediate-age pre-main sequence (\prems) stars (ages of $\sim5-30$ Myr) with
established distances and ages are rare in the astronomical literature
(\cite{Herbig78}).  Low-mass stars ($0.1-3$ \msol) in this age range are
predominantly past their active classical T Tauri phase, and are
usually called weak-lined T Tauri stars (WTTs) or post-T Tauri stars
(\cite{Martin97}).  Nearby stars ($d < 150$ pc) in this age range are
especially important for studies of stellar angular momentum evolution, stellar
multiplicity, the evolution of young and luminous brown dwarfs and planets, 
and the evolution
and longevity of circumstellar disks (e.g.\, \cite{Beckwith96},
\cite{Bouvier97},
\cite{Brandner98}, \cite{Low99}, \cite{RayJay99}, \cite{Bejar99}).  
It is particularly valuable to studies of these issues to
have samples of nearby, coeval, codistant stars in this age range.

Coronal X-ray emission (or more precisely, the ratio $L_x/L_{bol}$) 
is elevated
$1-3$ orders of magnitude above main sequence levels in low-mass stars
throughout the \prems\, phase (e.g., \cite{Briceno97}).
Pointed X-ray observations of nearby active star-forming molecular
clouds, where the bulk of the stars have modeled ages of $<$2 Myr,
have identified many new weak-lined T Tauri stars missed by previous surveys
(in Chamaeleon I (\cite{Feigelson93}), Taurus-Auriga 
(e.g. \cite{Strom94}), and other active star-forming regions).
Copious numbers of older \prems\, stars have been
found in star clusters such as the $30-50$ Myr-old clusters
IC 2602 and IC 2391 (\cite{Randich95}, \cite{Stauffer97}), around the
Orion molecular clouds ($<$7 Myr, \cite{Alcala98}) and associated with Gould's
Belt ($<$30 Myr, \cite{Guillout98}). Hundreds of isolated \prems\, 
and ZAMS stars have been discovered with X-ray telescopes, particularly the $ROSAT$ 
All-Sky
Survey (RASS, see reviews by \cite{Neuhauser99} and \cite{Feigelson99}), 
and thus provides an excellent means for locating older WTT stars 
which may no longer be proximate to their parent molecular cloud. 
Two stellar associations
with X-ray-selected samples of WTTs with ages around $5-20$ Myr lie
nearby: the Sco OB2 (Sco-Cen) association at $d \simeq 110-150$ pc
(e.g., \cite{deZeeuw99}, \cite{Preibisch99}) and the TW Hya T
Association at $d \simeq 50$ pc (\cite{Webb99}).

Last year, a new, nearby stellar aggregate was added to the list:  the
$\eta$ Chamaeleontis cluster (\cite{Mamajek99}, hereafter Paper I) with
age $t \approx 8$ Myr. The new cluster has $d$ = 97\,pc, and contains 13
known members with masses $0.1-3$ \msol\, within a very small region (0.2
square degree) of sky. The cluster was discovered with a \rosat\/ High
Resolution Imager (HRI) pointing of a tight group of 4 RASS X-ray
sources previously established to be WTTs (Alcala et al. 1995, 
Covino et al. 1997). The
low-mass X-ray-discovered stars have the lithium and H$\alpha$ spectral
signatures of WTT stars and are clustered around several
intermediate-mass stars including the $V$ = 5.5 B8V star $\eta$ Cha,
and the $V$ = 6, A8V+A8V double-lined, eclipsing binary RS Cha.
Paper I announced the group as the fourth nearest open cluster to the
Sun and the second nearest grouping of \prems\, stars after the TW Hya
Association. 

Since then, three additional candidate groups in the 
Sun's neighborhood have been announced: the Carina-Vela 
group (a possible extension of Sco OB2; \cite{Makarov00}) 
the ``Tucanae Association'' (\cite{Zuckerman00}), a
new association of $\sim$30-Myr-old post-T Tauri stars in
Horologium (\cite{Torres00}), and a small
group of T Tauri stars associated with the isolated
MBM 12 cloud (\cite{Hearty00}). As with
the $\eta$ Cha cluster, these new groups will require further
study to piece together the recent star-formation history
of the solar neighborhood, as well as investigate
the dynamics of disintegrating star clusters..

The present paper discusses the properties and origins of the 
$\eta$ Cha cluster in detail.  The X-ray data for cluster members is
presented in Section 2.  Section 3 gives preliminary evidence
for the existence of additional cluster members in the core region.
Section 4 compares the $\eta$ Cha cluster to other open clusters.  In
Section 5, we investigate the kinematics and origins of the $\eta$ Cha
cluster with respect to other young stars in the 4th Galactic
quadrant.  We argue that the $\eta$ Cha cluster, the Sco-Cen 
OB association, the TW Hya Association, and a group of young stars 
near $\epsilon$ Cha, likely formed in or near the same giant molecular
cloud complex $5-15$ Myr ago.  Section
6 discusses the interstellar medium (ISM) in the direction of the
$\eta$ Cha cluster and strengthens the claim for a kinematic origin in
Sco-Cen. Our findings are summarized in Section 7. Appendix A details
the kinematics of the $\eta$ Cha cluster, the TW Hya association,
the $\epsilon$ Cha group, and the Sco-Cen association. Appendix B 
presents notes on individual $\eta$ Cha cluster members. 
Particular attention is paid to the eclipsing binary RS Cha, 
which we find to be consistent
with being two \prems\, A stars, but which has been 
modelled in previous literature as a non-coeval 
post-main sequence system.

\section{X-ray Observations} \label{xray}

\subsection{Source Identifications}

\begin{figure}
\begin{center}
\epsfxsize=8cm
\epsffile{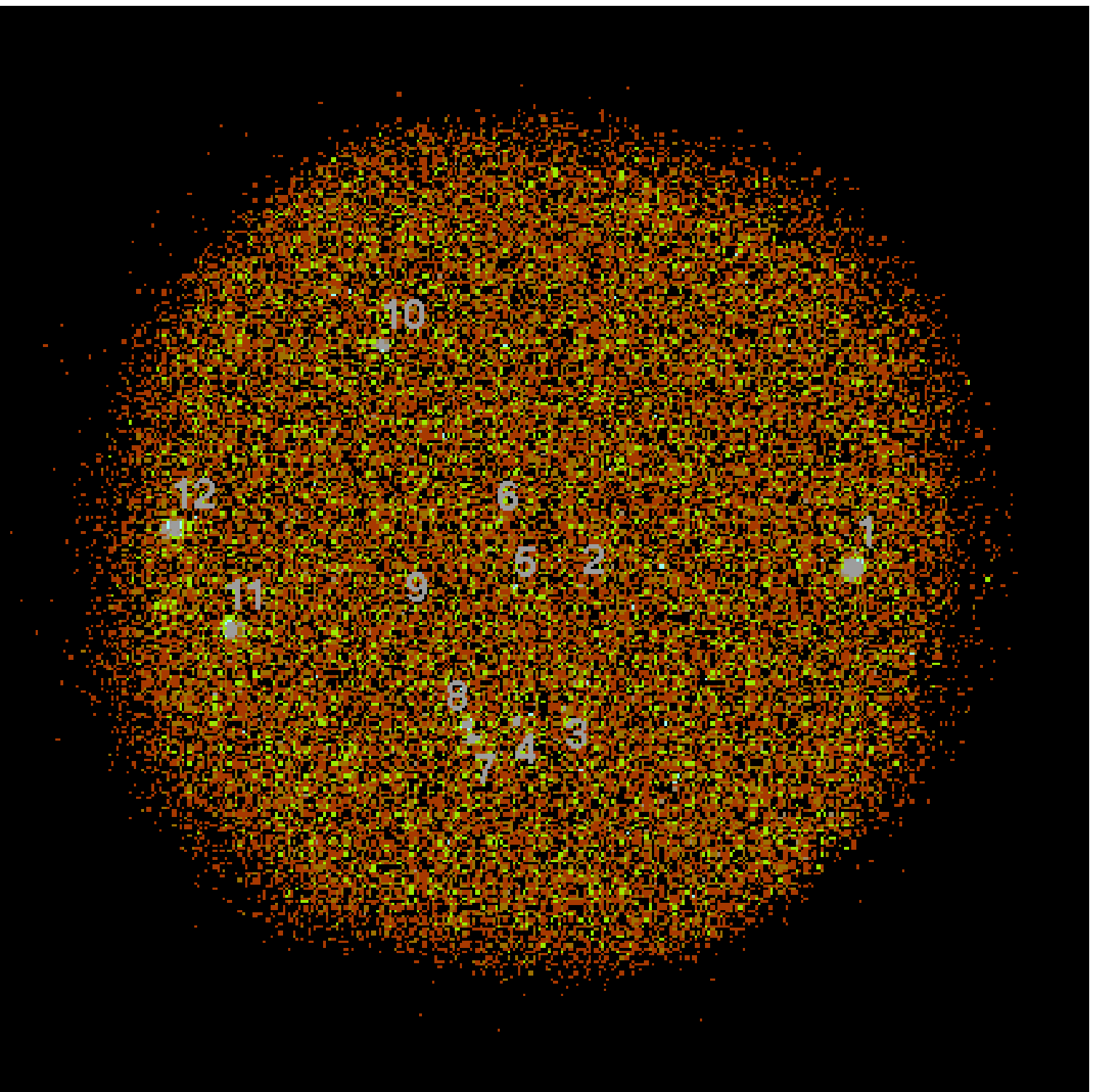}
\caption{\rosat\/ HRI image of the $\eta$ Chamaeleontis cluster.
The field is centered at $08^h42^m, -78^\circ 56\arcmin$ and 
has a diameter of 40\arcmin.  
\label{Fig_HRI}
}
\end{center}
\end{figure}

\placefigure{Fig_HRI}

We observed the $\eta$ Cha region with the HRI (\cite{zom95}) on board
the \rosat\/ satellite (\cite{tru83}) for about 12 hours in 32 segments
spread over several months in 1997.  Exposures were obtained during 4
satellite orbits over April $26-29$, 14 orbits distributed between
September 11 and October 8, and during a more concentrated group of 17
orbits in October $26-29$.  These data were analyzed within the
IRAF/PROS (version 2.5.1) software system (\cite{sao98}).  The image
received from the \rosat~Standard Analysis Software System (SASS
version 7) showed prominent sources, each showing identical structure
distributed over 15\arcsec\/ due to incorrect satellite aspect
solutions.  To alleviate this problem, the images were divided into
orbital intervals, centroids fitted to the brightest near-axis source,
and the images aligned and merged (\cite{Harris98}).  The reconstructed
image, shown in Figure \ref{Fig_HRI}, is improved with residual
distortions of a few arcseconds.  About 1.5 hours of data were lost in
this realignment process giving a net exposure of 41.7 ks.

Sources were located in the merged image with a cell detection
algorithm within the IRAF/PROS software system, where square cells with
sizes $6\arcsec \times 6\arcsec$ and $24\arcsec \times 24\arcsec$ are
passed along the image to find peaks exceeding signal-to-noise ratio
$S/N = 3$.  This ratio is computed using Poisson statistics after
subtraction of a local background level and was confirmed manually
using a global background level.  The centroid positions of each peak
was located with a likelihood ratio statistic.  Satellite boresight
aspect errors frequently translate \rosat~images by several
arcseconds.  This was corrected by aligning eight strong X-ray source
positions with their corresponding optical star positions, resulting in
a good fit with boresight residuals in a range $\pm 2\arcsec$.  With
the exception of RECX 12 where the right ascension is unreliable due to
off-axis telescope coma, the statistical X-ray positional uncertainties
are $1\arcsec-3\arcsec$ and the total uncertainties are about
$2\arcsec-4\arcsec$.

\placetable{Table_XRAY}

The resulting 12 X-ray sources and their properties derived from the
HRI data are listed in Table \ref{Table_XRAY}.  Following Paper I,
these sources are denoted by their \rosat~$\eta$ Chamaeleon X-ray
(RECX) number.  In columns 2 and 3, counts in the time-integrated image
are measured in circles ranging from $20\arcsec-57\arcsec$ diameter,
depending on the off-axis angle, after removal of a constant
background value of 0.0638 counts per square arc second.  Uncertainties
are based only on counting statistics.  The effective exposure time is
the detector live time of 41.7 ks corrected for location-dependent
detector quantum efficiency variations, telescope vignetting, and point
spread function scattering following SASS procedures.  This calculation
is not optimized to achieve the high signal-to-noise ratio; e.g., the
weakest source, RECX 9, shows $S/N = 2$ in Table \ref{Table_XRAY} but
has $S/N = 4$ when a smaller extraction circle is used.

In column 4, luminosities are estimated using a distance of 97 pc
(Paper I) and the conversion 1 count\,ks$^{-1}$
= $3 \times 10^{-14}$ erg s$^{-1}$ cm$^{-2}$ in the $0.1-2$ keV band.
Derived from a convolution of Raymond-Smith thermal plasma source 
spectra with the telescope and instrument response functions using 
the W3PIMMS software, this conversion is accurate to $\pm 0.2$ in
log \lx for any combination of source temperatures $kT > 0.2$ keV and
column densities $N_{\rm H} < 1 \times 10^{20}$ cm$^{-2}$.  Note that 
this \lx applies only to the $0.1-2$ keV band at the source, and not the
luminosity emitted over all energies which can be substantially larger.  

Stellar counterparts (column 5) were sought from the SIMBAD database,
Hubble Space Telescope Guide Star Catalog (version 1.2), United States
Naval Observatory PMM catalog (USNO-A2.0), and the Digitized Sky
Survey (DSS).  The offset between the X-ray source and stellar
counterpart in column 6 is based on optical positions given in Paper I.
The final columns summarizes X-ray variability characteristics of the
sources (see below), and USNO \bj\, and $F$ photometry (see \S 3.2
for discussion on USNO magnitudes)

In addition to these $S/N >3$ sources, potential weak sources with
lower $S/N$ or found only in a limited pulse height channel range (PHA
= $3-8$) where background rates are lower \cite{dav97} were examined on
both the HRI image and DSS.  One additional source was found: a marginal 
X-ray source at the extreme northern edge of the image near the star 
GSC 9398\_0105 with \bj\,=\,12.3, $F$\,=\,11.6. For its blue color, 
GSC 9398\_0105 is a few magnitudes too faint to be a member of the 
cluster, and we do not assign it an RECX number.

\subsection{X-ray spectra and variability}

The HRI has a limited spectral response capable of providing a single
hardness ratio within the $0.1-2$ keV band (\cite{pre98}).
Due to spatial variations in the detector response and changes in the
detector high voltage level made between segments of our observation,
we do not attempt a quantitative analysis.  But qualitative examination
of the pulse height data shows a distinct trend.  The brightest source,
RECX 1, is also the hardest.  Sources with intermediate luminosities 
around log \lx $\simeq 30.0$ erg s$^{-1}$, have intermediate hardness 
ratios consistent with thermal emission in the range $0.2 < kT < 1$ keV.
The fainter sources with log \lx $\simeq 29.0$ erg s$^{-1}$ have the 
softest spectra.  This trend is seen in T Tauri and other young star 
samples, and can be explained by relatively simple models of plasma 
heated by magnetic reconnection in loops near the stellar surface 
(\cite{pre97}).  

As the X-ray emission from young late-type stars is known to be highly
variable due to flares and other manifestations of magnetic activity,
the photon arrival times were examined.  Column 7 of Table
\ref{Table_XRAY} summarises the presence and degree of variability.
``Yes'' indicates that the source is variable at a $>$99\% confidence
level using nonparametric one-sample Kolmogorov-Smirnov and von Mises
tests a\-gains\t the hypothesis of a constant source, ``Possible''
indicates 90\%$-$99\% probability from one or both tests, and ``No''
indicates consistency with a constant count rate.  The approximate
minimum and maximum levels in HRI counts ks$^{-1}$ seen in the light
curves appear in parentheses.

\begin{figure}
\begin{center}
\epsfxsize=8cm
\epsffile{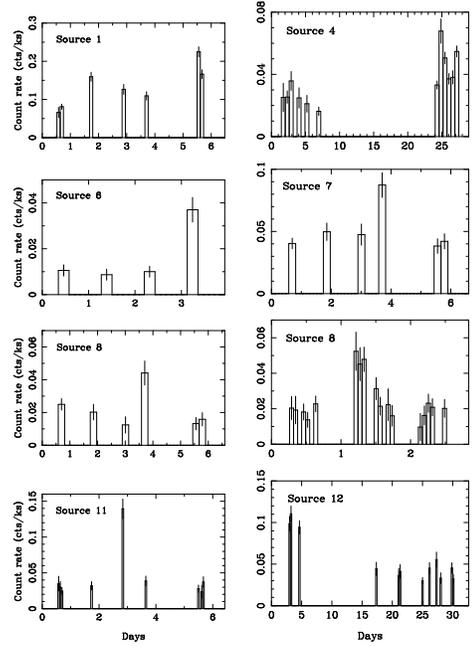}
\caption{
Selected X-ray light curves of highly-variable $\eta$
Chamaeleontis stars.  The source numbers refer to RECX source
numbers.  Starting times and bin sizes differ from panel
to panel.  For each source, starting times are: Source 1
(1997 October 3.2), 4 (October 3.2), 6 (April 26.5),
7 (October 3.2), 8 left (October 3.2), 8 right (October
26.5), 11 (October 3.2), 12 (September 8.9). 
\label{Fig_XRAY}
}
\end{center}
\end{figure}

\placefigure{Fig_XRAY}

Figure \ref{Fig_XRAY} shows the more dramatic changes seen in the 
seven variable sources.  Factors of $2-4$ jumps of flux on timescales 
ranging from 0.5 to $>$2 days are seen.  Fluxes often exhibit a 
factor of $3-7$ range over the entire observation with peak luminosities
between $10^{30.1-30.9}$ erg s$^{-1}$.  The light curves give the 
impression of individual powerful flares superposed on more moderately 
variable emission.  The best-recorded event occurred in RECX 8 (= RS Cha)
and shows a decay timescale of $\sim$0.5 days.

\subsection{Anomalous X-ray luminosity function}

While the X-ray properties of the $\eta$ Cha stars are similar in most
respects to those seen in other T Tauri populations
(\cite{Feigelson99}), the distribution of X-ray luminosities is quite
unusual: there are too few stars with lower, but detectable, X-ray
luminosities given the number of stars with high luminosities.  The
$\eta$ Cha luminosity function is roughly flat between $28.5 < \log 
L_x < 30.5$ erg s$^{-1}$, whereas source numbers are found to rise
inversely with log \lx in well-studied \prems\, and ZAMS 
stellar clusters.  Consider, for comparison, the Chamaeleon I
young stellar cluster (\cite{Lawson96}), $\alpha$ Per
(\cite{Randich96}) and the Pleiades (\cite{Stauffer95}) clusters.  In
these clusters, the number of X-ray sources in the range $28.5 < \log
L_x < 29.5$ erg s$^{-1}$ in the \rosat\/ band is about twice the number
of luminous sources with $\log L_x > 29.5$ erg s$^{-1}$.  The total
stellar population, including stars undetected in X-rays, is at least 4
times the $\log L_x > 29.5$ erg s$^{-1}$ subpopulation.  (This value
applies for the Pleiades, where the membership catalog is most complete
but still is likely missing some low-mass M and L dwarfs.)

Along an isochrone, luminosity scales with mass, and if the X-ray
emission is saturated (\lx/\lbol\, $\approx$\, 10$^{-3}$), then \lx\,
should scale with mass. Indeed, Feigelson et al. (1993) reported 
a correlation between $L_x$ and stellar mass in a \prems\,
population such that most $\log L_x < 29.0$ erg s$^{-1}$ stars 
have $M < 0.4$ \msol. Extrapolating from the number of $\eta$ Cha 
cluster stars with $\log L_x > 29.5$ erg s$^{-1}$
discovered in Paper I, a normal X-ray luminosity function predicts 
roughly 12 additional stars with $28.5 < \log L_x < 29.5$ erg s$^{-1}$ 
and about 16 stars with $\log L_x < 28.5$ erg s$^{-1}$. Our
HRI survey was sensitive to stars at the distance of $\eta$ 
Cha with $\log L_x > 28.5$. We surmise
that a considerable number of low-mass stars with detectable
X-ray luminosities reside outside the cluster core region surveyed
by {\it ROSAT} HRI.

Low-mass members of the cluster may have escaped
after the dispersal of the proto-cluster molecular cloud. 
The molecular cloud core that formed the cluster stars was most
likely dispersed by an external supernova or stellar winds,
or possibly by the outflows and winds of the most massive cluster members. 
The discussion in \S 6 elaborates on what is known about the interstellar
medium in the cluster vicinity, and points to the likely
culprit being the massive stars in the Sco-Cen OB Association.

If the cluster stars formed with a small velocity dispersion of 
$\approx$1 \kms (as found in the Cha I molecular cloud core; \cite{Dubath95}),
and the bulk of gas mass was dispersed, then the thermal motions
of the stars will lead to the slow evaporation of the cluster. The
escape velocity of the cluster, from summing the modeled masses
of the 13 known systems within 15\arcmin (0.4 pc) radius of the center,
is only $\simeq 0.5$\,\kms\,(Paper I). Doubling the cluster mass,
by finding unseen companions and possible X-ray faint\,/\,low-mass 
stars (see \S 3.1), would only increase 
\vesc\, by $\sqrt{2}$ to $\simeq$\,0.7 \kms. Numerical studies
(\cite{Lada84}) show that after a bound, embedded cluster
(N\,=\,50--100) sheds its molecular cloud, the cluster can lose 
10-80\% of its stars, depending on the time scale for gas dispersal
and the initial star-formation efficiency. If this scenario is 
correct, a ``halo'' of low-mass members  
within a few degrees of the $\eta$ Cha cluster core
(1$^{\circ}$ = 1.7\,pc at $d$\,=\,97\,pc), should be present.

\section{The Stellar Population of the $\eta$ Cha cluster}

\subsection{Initial mass function}

In order to estimate the total population of the $\eta$ Cha
cluster, we extrapolate a typical IMF assuming that we have
a complete census of stars with masses above a certain mass threshold.
We treat the membership of stellar primaries
with masses $>$1\,\msol\, as complete since the stars would either be in 
the magnitude-limited {\it Hipparcos} or Tycho-2 catalogs. 
The Tycho-2 catalog is 99\% complete to $V\,\simeq\, 11.0$ (\cite{Hog00}), 
which is 0.5 mag fainter than the $\approx$1.0\,\msol\, K4 star RECX 1.
The HRI field has
$5$ primaries with model masses $>1$\,\msol\, ($\eta$ Cha,
RS Cha, HD 75505, RECX 1 and 7, see Table 1 of Paper I), of which 
only RECX 7 has no astrometric catalog entry, 
likely due to its proximity to HD 75505.

\begin{figure}
\begin{center}
\epsfxsize=8cm
\epsffile{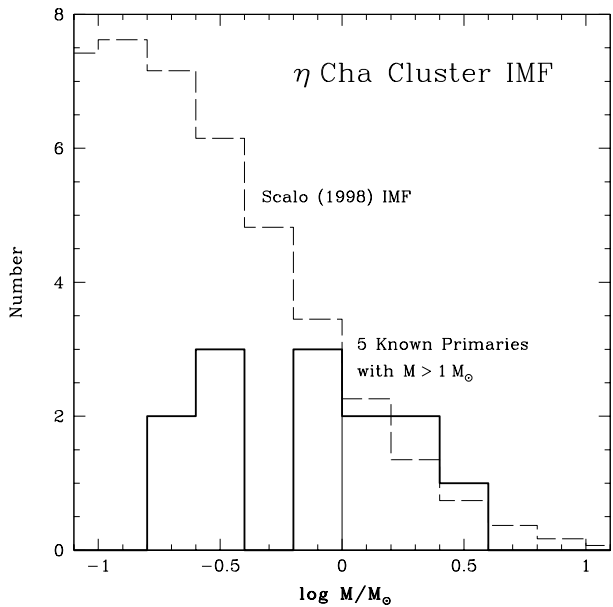}
\caption{
Mass function for the 12 X-ray stars and HD 75505 (thick line),
compared to the \cite{Scalo98} initial mass function (thin line).
The IMF is normalized to have five {\it primaries} above 1 
\msol\, (filled $\triangle$). A total membership
of 31\,$\pm$\,14 primaries is predicted, of which
13 have been discovered (see \S 3.1).
\label{Fig_IMF}
}
\end{center}
\end{figure}

A coarse extrapolation of a \cite{Scalo98} field mass function for a 
population with $5\,\pm\,\sqrt{5}$ systems with $M >$1\,\msol\, 
predicts a total membership of $\sim31\,\pm\,14$ stellar primaries with masses 
greater than $10^{-1.1}$\, =\, 0.08 \msol, the cannonical minimum 
stellar mass limit. Figure \ref{Fig_IMF} displays a histogram of the 
masses of the known cluster stars, along with an extrapolated 
Scalo mass function scaled to 5 $M >$1\,\msol\,stars.
The multiplicity statistics for the $\eta$ Cha cluster stars
await future study. Assuming 1.5 stars per primary, typical for T Tauri 
stars in the Taurus clouds (\cite{Kohler98}) and young main sequence 
stars in the Hyades (\cite{Patience98}), our IMF extrapolation suggests 
a total population of $\sim 50$ stars. Our current census includes 
12 X-ray detected stars and the common proper 
motion A star HD 75505, so we infer that perhaps $18\,\pm\,14$ 
additional {\it primaries} (i.e. systems) await discovery.
 
This extrapolation implies that our X-ray survey alone likely detected 
$39^{+32}_{-12}\%$ 
of the systems in the $\eta$ Cha core region.
Did our X-ray survey miss a large number of ``X-ray faint'' WTTs, or
have some of the cluster members dispersed outside of the field of 
view of {\it ROSAT} HRI field of view? Our discussion of the
anomalous X-ray luminosity function in \S 2.3 suggests that
many low-mass stars with detectable X-ray emission are likely
to exist outside of the HRI field of view. In the next section,
we also show preliminary evidence that stars with magnitudes comparable
to the WTTs discovered in Paper I also exist in the core region, but
were missed with HRI.

\placefigure{Fig_IMF}

\subsection{Clustered X-ray faint stars}

To test how complete the X-ray survey was, 
we examined the location of all stars selected through
photometric properties.  The X-ray and photometric selection have
been effective complementary approaches in other \prems\, membership
studies (e.g. \cite{fla99}).

Due to their proximity and youth, low-mass cluster members should
appear magnitudes above the vast majority of field stars in a
color-magnitude diagram (CMD).  At the time of writing we were 
lacking accurate optical magnitudes for the cluster members, however
new absolute and differential photometry (for rotation studies
and HR diagram placement)
will be presented in Lawson, Crause, Mamajek, \& Feigelson
(in preparation). For this study, we elected to 
use the USNO-A2.0 catalog (\cite{Monet98}) generated from the 
ESO/UK Schmidt photographic plates of the southern sky, which gives 
two-color photographic magnitudes for objects with $F < 18$.  The 
photographic $F$ magnitude is from the ESO-R IIIa-F plates, and is 
similar to Cousins $R$ (\cite{Lawson96}). The 
USNO \bj\, magnitudes from the UK-SRC IIIa-J plates
corresponds closely to Johnson $B$.  
Using data provided by Arne Henden of USNO, we derived the
following conversions:  Johnson \BV\, = 0.863\,\jminusf + 0.005 and 
Johnson $V$ = $F-0.5313$\,\jminusf $-$ 0.015. The
\jminusf~colors were redder than what the RECX spectral types (Paper I) 
would predict, and the photographic magnitudes are coarse enough that
we do not convert to the Johnson colors in our discussion. 
Although tied to the 
Tycho {\it BV\,} photometry (\cite{ESA97}), the USNO-A2.0 catalog is 
estimated to have absolute photometric uncertainties of $< 0.5$ mags 
and relative uncertainties around $0.15$ mag within a plate in the 
southern hemisphere.  We thus recognize that the CM diagram based 
on USNO magnitudes may not be a very precise tool for cluster 
membership study.

\begin{figure}
\begin{center}
\epsfxsize=8cm
\epsffile{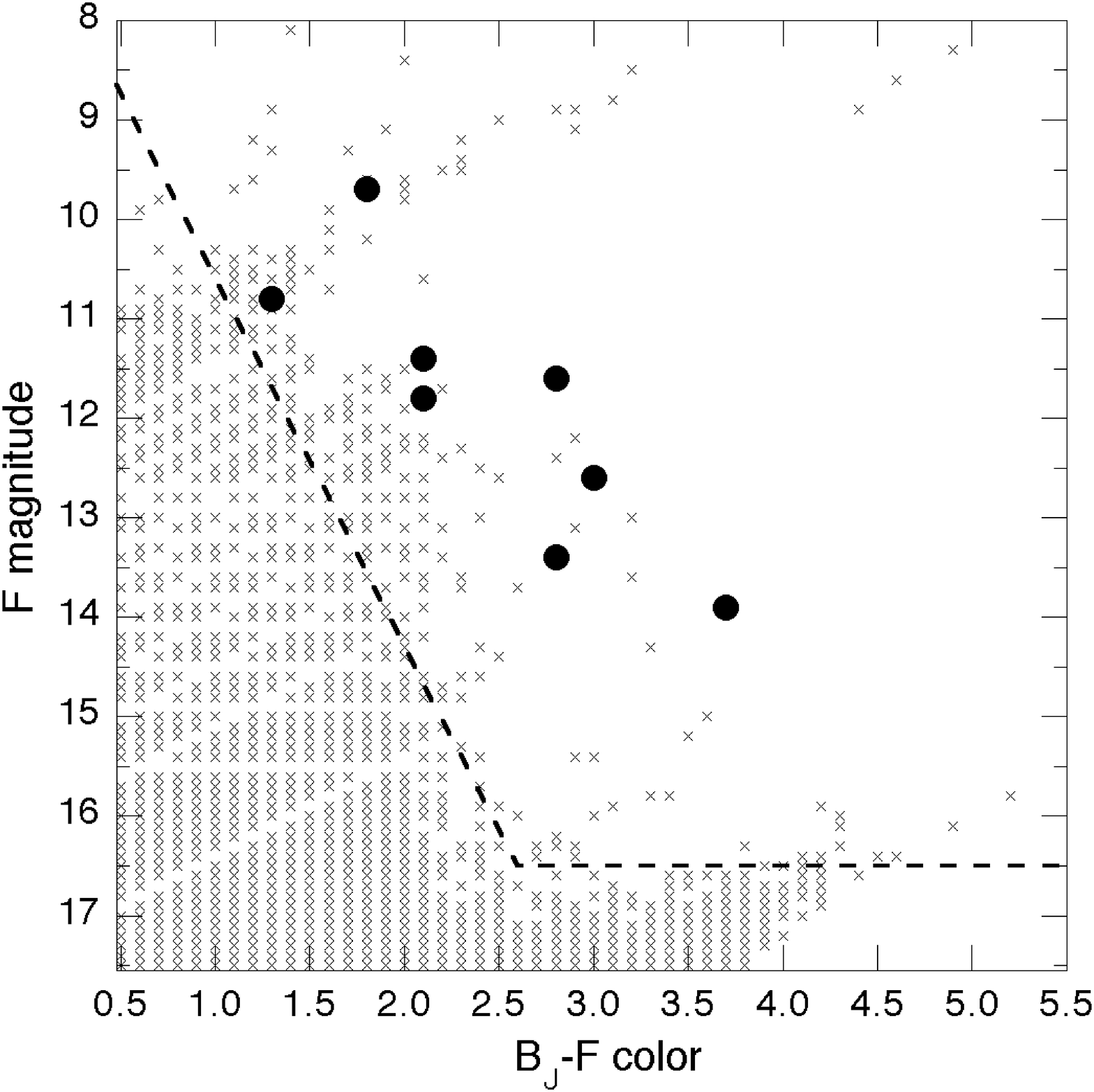}
\caption{
A color-magnitude diagram for 21,413 USNO A2.0 stars within a 
1$^{\circ}$ radius of 8$^{\rm h}$ 42$^{\rm m}$, $-79^{\circ}$.   
The $\eta$ Cha WTT stars are shown as large filled circles.  
The 351 stars selected for study in \S 3.2 lie above the 
dashed line.
\label{Fig_CMD}
}
\end{center}
\end{figure}

\begin{figure}
\begin{center}
\epsfxsize=6cm
\epsffile{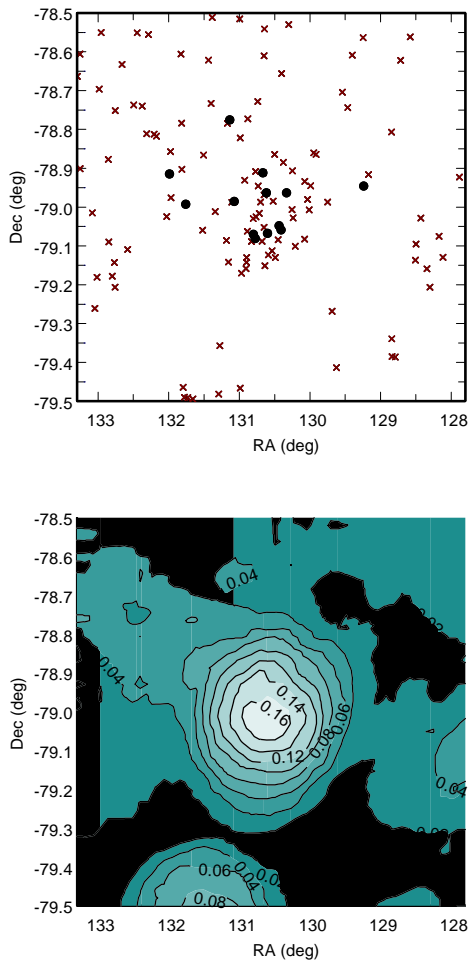}
\caption{
(Upper) Positions of the USNO photometric candidates (small
crosses) and known $\eta$ Cha cluster members (filled circles) in a
1 square degree box.
(Lower) A density map (stars arcmin$^{-2}$) of the
these stars smoothed with a 20\arcmin\, diameter kernel. 
\label{Fig_SDE}
}
\end{center}
\end{figure}

A 1$^{\circ}$ radius around $(\alpha,\delta) = (8^h42^m, -79^{\circ}
0^\prime)$ contains 45,058 USNO-A2.0 catalog stars. The
\jminusf--$F$ CM diagram for 19,747 stars with $F < 17.5$ and
$0.5 <$ \jminusf $< 5.5$ is shown in Figure \ref{Fig_CMD}, where the
X-ray selected $\eta$ Cha members are denoted with black circles. 
To isolate stars
with $F$ magnitudes not more than $\sim 2.5$ magnitudes fainter than
the X-ray stars but with similar colors, we identified 351 stars with
$F$ = 3.71$\times$\,\jminusf\,+\,6.86 and $F < 16.5$ as indicated by
the dashed lines.  The diagonal line approximates the location of the
ZAMS at 97 pc, and the $F = 16.5$ limit eliminates the thousands of
background stars in the CM diagram.  Note that the USNO magnitudes
are given in 0.1 magnitude steps, hence grid-like appearance of the
CMD. The celestial positions for these 
351 bright red stars exhibits a clustering of tens of objects above 
the average stellar density within $20\arcmin$\, of the center, both 
in the scatter plot and in the smoothed stellar density distribution
shown in Fig. \ref{Fig_SDE}. 

\placefigure{Fig_CMD}

\placetable{Table_USNO}

\placefigure{Fig_SDE}

To test whether the enhancement of starcounts is real\footnote{The
clustering of the photometric candidates is unlikely to be due to
extinction, as both $\eta$ Cha and RS Cha have $E(B-V) \simeq$ 0.00
(\cite{Westin85}, \cite{Clausen80}) and the wider region is nearly
dust-free (Section 5).}, simulations of random star positions were
constructed and smoothed with various Gaussian kernels.  We find that
the observed enhancement is present at the $>3\sigma$ confidence level
compared to the randomized samples.  The observed variation in stellar
density across the entire 1$^{\circ}$ field is 81\% (1-$\sigma$) of the
average density, compared to an average of 31\% (1-$\sigma$) across 5 
random fields.  The ratio of maximum/minimum density measured in the 
field was 56, compared to 11 $\pm$ 5 for the synthetic datasets.  
Examination of stars fainter than $F = 16.5$ and to the left of the 
diagonal line in Figure \ref{Fig_CMD} showed no statistically significant 
density enhancements compared to synthetic random fields.

Based on the background stellar density, we find $\sim 50$
photometric candidates in the core region (the size of the HRI
field; 40\arcmin), of which a fraction will likely be confirmed 
as new, low-mass members.  The youth of the candidates will be tested 
when spectroscopy and proper motion studies are completed. 
The photometric study by Lawson et al. (in preparation) will
address the issue of how many of the USNO candidates in the cluster
core region are legitimate members. New WTTs
discovered in the region covered by the HRI survey will either be (1) 
X-ray faint (\lx\,$<$\,10$^{28.5}$), most likely the least-luminous
and lowest-mass cluster members or (2) as companions to previously
known members.

\section{Cluster properties and comparison with other open clusters}

Over 1100 open clusters are known and many have well-characterized
properties such as distance, age, mass, size, concentration,
metallicity and so forth (\cite{lyn87}; \cite{Mermilliod95}).  Table
\ref{Table_LYNGA} summarizes the properties of the $\eta$ Cha cluster
for comparison with other clusters.  The official designation format
follows that of the 5th Catalogue of Open Clusters (\cite{lyn87})
updated to J2000 as recommended by Lortet et al.\ (1994).  The
cluster center location is estimated here from the smoothed stellar
density shown in Figure \ref{Fig_SDE}, and agrees with the average
position of the known members to within $1\arcmin$.

The brightest star characteristics are obtained from the {\it Hipparcos\,}
catalog (\cite{ESA97}).  The extinction is derived from the Stromgren photometric excess 
{\it E\,}($b-y$) = --0.004 reported for $\eta$ Cha (\cite{Westin85}).  The spectral 
type and \BV~color of HD 75505 imply 0.4 mags of extinction,
however this is likely due to spectral misclassification (see Appendix).

The angular diameter is estimated from top panel of Figure \ref{Fig_SDE}; 
it may be underestimated if a halo of cluster members surrounds a denser 
core.  The cluster age is from Paper I.  The membership population is 
estimated from \S 3.1 and 3.2, and the $>$ indicates that additional
members may exist that have either (1) low masses and luminosities, 
(2) are X-ray faint or (3) are outside the HRI field-of-view.  
The cluster mass is estimated to be $\approx$23 \msol\, from the 
13 known members ($\approx$\,13\,\msol) characterized in 
Paper I and Appendix B, 
plus a $12\,\pm\,5\,$\msol\,contribution from $34\,\pm\,14$ missing members 
that should exist (\S 3.1) with assumed average masses of 0.35\,\msol 
(the average mass of the stars below 
1\,\msol\, in the extrapolated IMF shown in Fig. 3).  
Trumpler classifications were derived
by us following the prescription of \cite{Trumpler30}. The notation
means the cluster has: detached, slight concentration (II concentration
class), bright and faint stars (3 range of brightness class), and a
poor population (p richness class).  The radial velocity is taken to be
the weighted mean of RS Cha (\cite{Andersen75}), RECX 1, RECX 10, and RECX 12
(\cite{Covino97}),
while $\eta$ Cha (variable $v_r$; \cite{BB2000}) and RECX 7 (discordent single
measurement; Covino et al. 1997)
were omitted (see also \S 5.1).
Galactic location and heliocentric velocity vector 
is reproduced from \S 5.
The cluster proper motion is the weighted mean of
the individual proper motions of $\eta$ Cha, RS Cha,
HD 75505, RECX 1, and RECX 11.

Comparing to the large sample of other well-char\-acterized open
clusters (\cite{Lynga82}, \cite{Janes88}), the $\eta$ Cha cluster has a
typical central star density ($\sim 200$ \msol\,pc$^{-3}$)
and Z-distance from the Galactic plane for clusters of its age (Z = 36 pc).  
It is not
part of any of the three established complexes of young open clusters
(Perseus, Carina, Sagittarius) which are centered $1.4-2.3$ kpc away
from the Sun, but is likely associated with the Sco OB2 association 
(\S 5) which is not included in the open cluster lists.

Other properties are unusual compared to most open clusters.  The
brightest star is underluminous compared to other $\sim$10 Myr old
clusters. This is probably due to the poor membership of the $\eta$ Cha
cluster compared to more distant larger clusters.  The linear size of the
$\eta$ Cha cluster is also unusually small, which may be real or may
indicate that we have detected only the core of a larger structure.
But most striking is the proximity to the Sun: only three of $>1100$
open clusters are closer than $\eta$ Cha (Fig. \ref{Fig_LYNGA}).
While this is not an intrinsic property of the cluster, it makes
detailed study of the characteristics of its $\sim$10 Myr-old stars,
including very-low mass brown dwarfs, much easier to study than
in other similarly-aged clusters (\S 7.2).

Although the $\eta$ Cha cluster is smaller, and more poorly populated
than the average open cluster in the Lynga, sample, 
its population and youth
resemble that of the small clusters of \prems\, stars in nearby
molecular clouds like Chamaleon I, Lupus, etc. \cite{Gomez93} analyzed
the spatial distributions of T Tauri stars in several nearby molecular
clouds, and found that most are in clusters with memberships of tens,
with cluster radii of $0.5-1$\,pc. The $\eta$ Cha cluster has a similar
inferred population and size, except that it has no molecular gas 
associated with it (see \S 6). The $\eta$ Cha cluster therefore appears
to be what a cloud core with an associated YSO population 
would resemble several Myr {\it after} the main burst of star-formation, 
{\it after} the natal molecular gas has been dispersed, but {\it before} 
the cluster has disintegrated.

\placefigure{Fig_LYNGA}

\section{Kinematics and Origin in the Sco-Cen Association
 \label{seckinematics}}

In order to investigate the origin of the $\eta$ Cha cluster, we
study the kinematics of its members and of other young stars and
star-forming associations that may be related to it.  Paper I mentioned
that the $\eta$ Cha cluster appeared to share proper motions with the
large Lower-Centaurus Crux subgroup of the Sco OB2 (Sco-Cen)
association.  We consider this in detail here, and include other nearby
individual and groups of young stars to investigate possibilities of a
common origin. Besides the $\eta$ Cha cluster, we include the 
three primary subgroups of Sco-Cen, the nearby TW Hya association,
and the group of young stars near and including $\epsilon$ Cha
(details on each group are given in Appendix A).

Galactic motions and vectors were calculated using the algorithm of
\cite{Skuljan99}.  The Sun's current position
is placed at the origin, and the UVW velocities we present
are heliocentric. The algorithm factors in the parallax,
position, proper motion, and the 10 {\it Hipparcos\,} covariance
coefficients to produce a Galactic position and velocity with
propagated uncertainties. For known widely-separated multiple 
stars, we use the Tycho-2 catalog, which has a much longer
time baseline than the {\it Hipparcos} astrometry. 
When Tycho-2 proper motions (\cite{Hog00}) are used,
the covariance coefficients are set to zero. 

Our extrapolations of past stellar
motion assume linear ballistic trajectories.
Galactic differential rotation, 
the Z-direction gravitational potential, or possible gravitational 
deceleration by the parent molecular clouds were not included.
Positional uncertainties of past locations, 
typically a few parsecs for individual stars, are dominated by velocity 
uncertainties in projection, and are ignored.
These results are presented in Table \ref{Table_KINE}. 
Note that in our discussions, weighted means are weighted by
the inverse variances.

\begin{figure}
\begin{center}
\epsfxsize=8cm
\epsffile{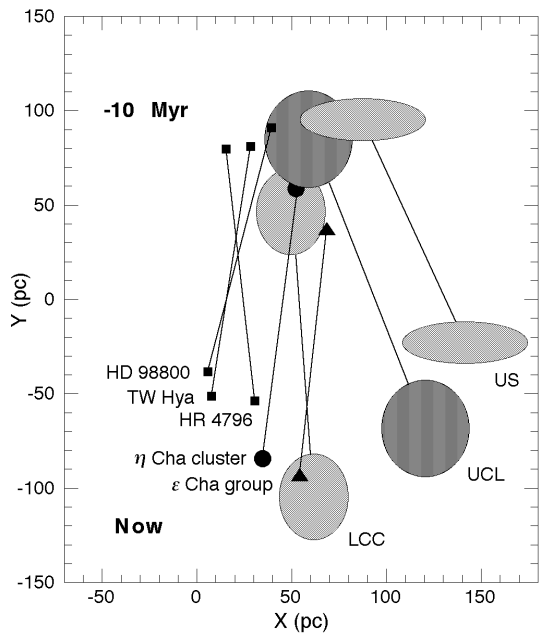}
\caption{
The Galactic positions of the $\eta$ Cha cluster, TW Hydra
association, the $\epsilon$ Cha group, and Sco-Cen OB subgroups 
now and projected back 10 Myr ago based on Table 3.  The motions
are with respect to the LSR; the solar peculiar motion 
(\cite{Dehnen98}) has been subtracted. The shapes of the Sco-Cen subgroups 
reflect their 3D extent, with the radii in X and Y being one standard 
deviation of the positions of all of the subgroup members as given by Z99. 
\label{Fig_XYZ}
}
\end{center}
\end{figure}

These kinematic data permit us to determine the locations of the
various stars and stellar associations in the past.  These results are
given in Table \ref{Table_KINE}.  Columns $9-11$ of Table
{\ref{Table_KINE} identify the Sco-Cen subgroup they were closest to in
the past, along with the time and approximate distance 
of closest approach ($\Delta$).  
Projected uncertainties in the past positions can be
estimated from multiplying the 1$\sigma$ uncertainties in the (U,V,W)
velocities times 10 Myr (1 \kms corresponds to 1.02 pc\,Myr$^{-1}$).
The extrapolated positional uncertainty radii 10 Myr in the past 
are $\approx$10\,pc for the $\eta$ Cha cluster, 20--30\,pc for 
the individual TW Hya members and $\approx$40\,pc for the TW Hya
association. The past positional uncertainty radii range from 15--50
pc for members of the $\epsilon$ Cha (the latter high value coming
from the motion of $\epsilon$ Cha itself). Owing to the large sample
sizes of stars with {\it Hipparcos} proper motions and parallaxes
(\cite{deZeeuw99}, Z99 hereafter), we conservatively estimate the 
positional uncertainty radii 
of the Sco--Cen subgroups to be $\leq$10\,pc. 

\placetable{Table_KINE}

\subsection{Results}

We find that the majority of these
stars were much closer together about $10-15$ Myr ago.  Fig.
\ref{Fig_XYZ} shows the projected current and past positions of the
major associations and clusters 10 Myr in the (X,Y) plane. The solar
peculiar motion of Dehnen \& Binney (1998) has been subtracted, so
that Fig. 6 is in the Local Standard of Rest (LSR). The $\eta$ Cha cluster 
was within $\Delta\sim$~13 pc of the centroid of the Sco-Cen LCC subgroup
11 Myr ago. The {\it group} motion
of the 4 TW Hya members suggest that it was within $\Delta \sim$~35\,pc
of the LCC 19 Myr ago. Similarly the {\it group} motion of the
$\epsilon$ Cha members places its centroid within $\Delta \sim$\,22\,pc of the 
LCC about 13 Myr ago.  The ages of the $\eta$ Cha, $\epsilon$ Cha,
and TW Hya members suggest that they, too, were formed $\sim$10 Myr ago
when they were much closer to the Sco-Cen association.
.
TW Hya, HD 98800, and HR 4796 are famous isolated young stars with ages
of $\approx$10 Myr (\cite{Webb99}), and their individual motions
indicate they were $\sim$20 pc away from Sco-Cen subgroups $\sim$15 Myr
ago. The kinematics and HR diagram position of TWA 9 suggest it is either 
not a member, or the {\it Hipparcos} distance is too small (see Appendix A.4).
Without TWA 9 factored in, the group motion of the other three TW Hya members
shows they were within 28 pc of the UCL subgroup 14 Myr ago, and 
within 21 pc of the LCC subgroup 21 Myr ago.
 
We suggest that many of the TW Hya association members may
have formed from a small cloud which was part of the Sco-Cen giant molecular
cloud complex which formed the subgroups, but had a slightly different 
motion. If the HR diagram ages are to be believed, then the bulk of the
star-formation in the proto-TW Hya cloud took place several Myr after the
cloud was closest to the Sco-Cen complex. The same pattern is
seen with the $\eta$ Cha cluster $-$ the HR diagram ages are several
Myr younger than the time of its closest pass to a Sco-Cen subgroup in the
past. 

The motions of the $\epsilon$ Cha group stars show that they, too, 
are moving away from Sco-Cen. Terranegra (1999) found an age of 
15 Myr for RX J1159.7-7601, and
we find that it was $\Delta\,\sim\,14\,$pc of the LCC group (age: 11--12 Myr)
about 16 Myr ago. We calculated a modeled age of HD 104036 of
6--7 Myr, and we find that it was $\Delta\,\sim\,17\,$pc of the LCC group 
about 7 Myr ago. It appears that many young stars in this region are 
moving away from Sco-Cen,
and their \prems\, model ages are comparable to the times when they 
were closest to Sco-Cen subgroups.

The motions of the Sco-Cen subgroups themselves show
an interesting pattern. Curiously, the minimum pass distances 
between the groups are roughly equal to the nuclear age of
the youngest of the two subgroups.
The two oldest subgroups, LCC and UCL, were closest about
11 Myr ago ($\Delta\sim$~47 pc versus $d\sim$~71 pc today).
The US and UCL subgroups were closest to each other $\sim$4 Myr ago 
($\Delta\sim$~46 pc versus $d\sim$~53 pc today).
Note that the ages derived from Walraven photometry for the
subgroups are $5-6$ Myr for US, $14-15$ Myr for UCL, and $11-12$ Myr
for LCC (\cite{deg89}). The distances between all of the subgroup 
centroids are currently increasing. 
The 3 major subgroups of the Sco-Cen association
were in a more compact configuration during their
bursts of star-formation $5-15$ Myr ago. 

\placefigure{Fig_XYZ}

The times and ``impact-parameters'' for 
individual stars and the Sco-Cen subgroups have some spread, however 
it is clear that the groups were much closer to each other
at a time when the two oldest subgroups were undergoing their
bursts of star-formation. None of the groups are currently moving
toward each other, and their motions appear more-or-less
radial, as if they were all associated with the giant molecular
cloud complex that formed the UCL and LCC subgroups 10 and 15 Myr
ago, respectively. 

\section{The interstellar medium around $\eta$ Cha  \label{secism}}

Due to the youth of the $\eta$ Cha cluster, it is possible the stars
were formed in contemporary molecular clouds although it has been 
recognized that its immediate vicinity has no large clouds
(\cite{Alcala95}, \cite{Feigelson96}, \cite{Mizuno99}). The Chamaeleon
I cloud that has been actively forming stars for at least several
million years (\cite{Lawson96}) lies 8$^{\circ}$ to the east, or
$\approx$70\,pc away three-dimensionally.  The
sensitive $^{13}$CO survey of the entire Chamaeleon region by Mizuno et
al. (1999) located two dozen very low-mass cloudlets around the three
primary Chamaeleon molecular clouds (Cha I, II and III), each with
$\sim 10$ \msol\, and a total mass of 500 \msol\, in molecular
material.  But none are closer than $\simeq 2^\circ$ to $\eta$ Cha:
there are no $>$ 2 \msol\, clouds of gas with densities $\geq$10$^{3}$
cm$^{-3}$ within several parsecs of the $\eta$ Cha cluster core.

There are, however, large-scale structures in the area that may be
relevant to the origin of both the Sco-Cen association and the $\eta$
Cha cluster.  A large H {\small I} filament runs parallel to the
Galactic plane at $b \simeq -25^{\circ}$, first discovered as a ``weak
ridge'' in an early 21-cm southern survey by McGee \& Murray (1961).
The feature was suggested to be the southern counterpart of the
well-studied neutral hydrogen Loop I (or North Polar Spur) by Fejer \&
Wesselius (1973).  This ``Southern Hemisphere Low Velocity Filament''
has a mass of order 10$^4$ \msol\, and a distance around
$100-115$\,pc estimated from stellar polarization studies
(\cite{cle79}, \cite{mor81}), similar to the {\it Hipparcos}
distance to Lower Centaurus Crux (118\,pc; Z99).

\placefigure{Fig_IRAS}

The H {\small I} filament is clearly related to a long arc of cold dust
in \iras~far-infrared maps, as shown in Figure \ref{Fig_IRAS}.  If one
interprets it as a portion of a spherical shell, the center lies
around  $(\alpha,\delta)=(12^h,-63^{\circ})$ or
$(\ell,b)=(297^{\circ},-1^{\circ})$ in the middle of the current
location of the Sco-Cen LCC subgroup. de Geus (1992) discusses this in
the context of various interstellar shells and loops that surround the
Sco-Cen OB association.  The inner ``wall'' of the Loop I bubble
has been detected as an ISM density jump $40\,\pm\,25$ pc away 
in the direction of Sco-Cen (\cite{Centurion91}). Putting
these distances into a model of a spherical bubble centered on 
Sco-Cen places the $\eta$ Cha cluster within the Loop I
bubble.  The morphologies and distance estimates to
these structure strongly suggest that past OB winds and supernova
remnants from the most massive Sco-Cen stars evacuated a large volume
within which the $\eta$ Cha cluster now lies.  This evacuation explains
why there is very little molecular material currently present near
$\eta$ Cha, TW Hya or the other stars likely originating in the Sco-Cen
giant molecular cloud.  The cloudlets found by Mizuno et al.\ (1999)
may be molecular material that has not been fully dispersed or
evaporated within the hot Sco-Cen supershell.

\section{Discussion}

\subsection{Summary}

The principal results of this study are:

(1) We have convincingly established that the group of X-ray 
selected stars around $\eta$ Cha constitutes a physical open cluster, 
and not a chance superposition of unrelated stars.  In addition to the 
arguments in Paper I (spatial coincidence of high- and low-mass 
stars; identical distances and motions of the brighter stars from 
{\it Hipparcos\,} measurements, and a self-consistent HR diagram), 
we find a compact cluster of photometric candidates coincident with
the X-ray-discovered population (\S 3, Figure \ref{Fig_SDE}).
The $\eta$ Cha cluster is smaller and more poorly populated than
most classical open clusters (\S 4, Table \ref{Table_LYNGA}). Its
size and population is comparable to the small clusters
seen in nearby stellar nursuries (e.g. Taurus, Lupus, etc.),
but the model ages are older ($\approx$8 Myr) and there is 
no associated molecular gas or dust.

(2) The X-ray selected stars exhibit very high levels of magnetic
activity, with powerful and high-amplitude X-ray variability (Table
\ref{Table_XRAY}, Figure \ref{Fig_XRAY}).  This is supported by
photometric evidence for large star\-spots on the stellar surfaces
(Lawson et al., in preparation). 

(3) The kinematics of the $\eta$ Cha and other young stellar groupings
over a large region of the southern sky indicate that many have a
common origin about 10-15 Myr ago during the star-formation epochs
of the Upper Centaurus Lupus and Lower Centaurus Crux subgroups of
the Sco-Cen OB Association (\S 5, Table \ref{Table_KINE}, Figure
\ref{Fig_XYZ}).  In particular, both the $\eta$ Cha cluster and TW Hya
association appear to be outliers of the $\approx$11-Myr-old 
LCC subgroup of the Sco-Cen association.  
We conclude that Sco-Cen is far larger than usually
assumed (e.g. \cite{deg89}, Z99), and that more young
stars or groups of stars will be found with motions consistent with
an origin in or near Sco-Cen.  

The presence of stellar groups like $\eta$ Cha and TW Hya lying 
today $\sim
50$ pc from the core of the Sco-Cen association can be readily
understood as the consequence of their velocities inherited from the
parent giant molecular cloud (\cite{Feigelson96}).  It is
well-established that, on large $50-100$ pc scales, giant molecular
clouds exhibit velocity gradients and dispersions around 5 \kms
(\cite{Larson81}, \cite{Myers83}, \cite{Efremov98}).  This high
velocity dispersion is usually interpreted as the consequence of
turbulence in cloud complexes. If the outlying stellar groups were
unbound from the main OB association, they would have dispersed at a
rate around 5 \kms for 10 Myr, which corresponds to their observed
$\sim 50$ pc separations from the association today.

The fate of the molecular material from which the $\eta$ Cha stars
formed is less certain.  One possibility, mentioned by Jones \&
Herbig (1979) in another context, is that moving clouds experience
resistance from the ambient medium and become separated from their
newly formed stars on timescales of $10^7$ years.  But the findings
reviewed in \S 6 suggest that the $\eta$ Cha cluster resided in a
changing interstellar environment.  Initially, the cloud from which
the cluster formed may have shared the stellar motion.  At some
later time the expanding stellar winds and/or supernova bubbles from
the most massive Sco-Cen stars caught up to the cluster and evacuated
or evaporated the $\eta$ Cha cloud material.  

As the H {\small I} and
dust filament is just a few degrees southwest of the cluster today
(Figure \ref{Fig_IRAS}), this cloud-stripping may have occurred quite
recently.  If the star-formation efficiency of the $\eta$ Cha cluster
progenitor cloud was around $5-20$\%, then the cloudlet was several
hundred solar masses comparable to the Cha I and II clouds which are
active today.  It remains to be investigated astrophysically whether
the Sco-Cen supernova remnants, now seen as the H {\small I} filament
with $M \sim 10^4$ \msol\, could disperse a molecular cloud of this
mass in the requisite time period.  It is possible that the tiny
molecular clouds found by Mizuno et al.\ (1999) throughout the
Chamaeleon region are left over from this dispersal process.

The kinematic tie between the $\eta$ Cha, $\epsilon$ Cha, and TW Hya
aggregates and the enormous Sco-Cen OB complex suggests that many of
the young stars in the fourth Galactic quadrant may have their origins
in the Sco-Cen giant molecular cloud.  The three young stellar groups
discussed here --- $\eta$ Cha, $\epsilon$ Cha, and TW Hya --- may be only
a small fraction of a large population of $\sim 10$ Myr \prems\, stars stars
distributed in a ``halo'' around the main concentration of the Sco-Cen
association. Some of these \prems\, stars will lie very close to the Sun.

\subsection{Future Work}

While study of the dynamical state and history of $\eta$ Cha is beyond
the scope of this paper, we can make some preliminary comments on this
topic.  The escape velocity from the cluster is likely to be about
$\simeq 0.5-0.7$ km s$^{-1}$ near the edge of the field that
{\it ROSAT} HRI surveyed. 
Molecular clouds with several hundred masses of gas, a likely progenitor
of the $\eta$ Cha cluster, have internal
three-dimensional velocity dispersions around $1-2$ km s$^{-1}$
(\cite{Larson81}).  If stars inherit the velocity dispersions of
their progenitor molecular clouds (\cite{Feigelson96}), then many stars
may have escaped from the cluster, and the cluster core we study
here may be evanescent.  

The dynamical history of the cluster is
probably more complex than a simple loss of high velocity stars, as
it depends on the rate of gas dissipation (e.g., slow thermal
evaporation or sudden removal by supernova remnants).  The role of
binaries and possibility of mass segregation also should be studied
(e.g. \cite{Mathieu85}, \cite{Bonnell98}).  Radial velocity
measurements of more cluster members and N-body simulations are
needed to address these questions with confidence.  It would also be
interesting to compare the dynamical evolution of $\eta$ Cha with that
of the TW Hya association, which probably had a similar origin in the
Sco-Cen giant molecular cloud but is now clearly unbound and
dispersed. Future kinematic simulations should also include
the effects of galactic differential rotation, and the gravitational
field of the disk in the Z direction.

The relative proximity of this cluster of stars, unobscured by
molecular material, makes the $\eta$ Cha cluster an ideal laboratory
for studying aspects of the evolution of intermediate-aged \prems\, stars
and substellar objects, such as angular momentum evolution along the
Hayashi tracks, stellar multiplicity, the luminous phases of brown
dwarfs, and the evolution and longevity of circumstellar disks.   At a
distance of 97 pc and an age of $\sim$10 Myr, the low-mass stellar
limit is defined by $I \approx 14$ and spectral type M5 -- M6.  Brown
dwarfs with masses $20-70$ $M_{Jupiter}$ are relatively bright and
hot objects with $I \approx 14-17$, $T_{\rm eff} = 2700-3100$ K, and
spectral types M6 -- M8.  These are more easily located and studied
than the under-luminous $T_{\rm eff} = 1500-2000$ K, L-type and T-type
objects that characterize older brown dwarf populations
(\cite{Kirkpatrick99}).  Identification of additional cluster members
can be provided by a combination of optical spectroscopy (spectral
typing and detection of key T Tauri-star diagnostic lines such as
H$\alpha$ and Li I $\lambda$6707), optical photometry and proper
motion studies.  The $\eta$ Cha cluster may even be a suitable location
for searching for the presence of early planet formation; current
$8-10$-m-class instruments already allow the resolution of sub-solar
system-sized structures at $d \approx$ 100 pc.

\acknowledgments

EEM thanks the J. William Fulbright program and the Australian-American
Education Foundation (AAEF) in Canberra for support.  
WAL's research is supported
by the Australian Research Council Small Grant Scheme and University
College Special Research Grants; EDF's research is supported in
part by NASA contract NAS8-38252 and NAG5-8422.  This research made use
of the SIMBAD database operated by the CDS in Strasbourg France, the
SkyView service provided by the HEASARC at NASA-GSFC, the ESA {\it
Hipparcos\,} and {\it Tycho\,} data\-bases, the USNO-A2.0 catalog, and
DENIS data obtained at the European Southern Observatory.  We thank
Jovan Skuljan of the University of Canterbury for graciously sharing 
his kinematics code in advance of publication. Thanks also go to Arne
Henden (USNO) for his discussions on photographic magnitudes.

\appendix

\section{Notes on the Kinematics of the Young Stars}

\subsection{Kinematics of the $\eta$ Cha Cluster}

Five members of the $\eta$ Cha cluster have proper motion data from
either the {\it Hipparcos}, Tycho-2, or STARNET catalogs:  $\eta$
Cha, RS Cha, HD 75505, RECX 1 and RECX 11. Only $\eta$ Cha and RS Cha
have well-determined parallaxes from the {\it Hipparcos\,} catalog,
whilst HD 75505 and RECX 1 have poor Tycho parallaxes.  Three of
these systems have published radial velocity ($v_r$) data: $\eta$~Cha
(\cite{BB2000}), RS Cha (\cite{Andersen75}), and RECX 1 
(Covino et al. 1997).
$\eta$ Cha has a variable $v_r$ and may be a binary; four measurements
have been made ranging from $-17$ to +37\,\kms (\cite{neu30,bus61})
with a mean of +14\,\kms (\cite{BB2000}). $\eta$ Cha (B8V) was detected in our
HRI survey, and we suspect the source of the radial velocity 
variations and variable X-ray emission is a low-mass companion.
 RECX 1 has $v_r$ = +18\,$\pm$\,2 \kms (\cite{Covino97}). The
eclipsing binary system RS Cha was carefully monitored by several
authors in order to refine the physical parameters of the system (see
Appendix A, section 7). A precise radial velocity for the system, $v_r = 
+15.9\,\pm\,0.5$ \kms, is given by Andersen (1975).  RS Cha is the only
system in the $\eta$ Cha cluster which has a well-determined $v_r$,
proper motion, and parallax producing a high-quality Galactic motion
vector of (U,V,W)$ = (-12.3, -19.1, -10.6)$ \kms with uncertainty
(\dudvdw)=(0.8, 0.5, 0.4)\,\kms.

Due to the quality of the parallaxes for both $\eta$ Cha and RS Cha, as
well as their location near the majority of cluster members WTTs, we
adopt their weighted mean parallax $<$$\pi$$>$ = 10.28\,$\pm$\,0.31
mas and Galactic position as the cluster center (X,Y,Z) =
$(34.6, -83.6, -35.9)$\,pc and (\dxdydz) = (1.1, 2.6, 1.1)\,pc. 
Using the weighted mean proper motions, radial velocities, 
and the mean $<$$\pi$$>$, we calculate the heliocentric cluster motion 
to be (U,V,W)$ = (-11.8, -19.1, -10.5)$\,\kms and
(\dudvdw)=(0.6, 0.5, 0.3) \kms. This translates into
a convergent point of ($\ell, b$) = ($211.7^{\circ}, -25.1^{\circ}$) 
with a total space motion of 24.8 \kms. 
This is close to the Gould's Belt convergent point 
found in the RASS-Tycho-2 sample by \cite{Makarov00}: 
($\ell, b$) = (244.3$^{\circ}$, --12.6$^{\circ}$), and 
Eggen's (1995) Local Association or ``Pleiades Supercluster'' 
($\ell, b$) = (240.6$^{\circ}$, --27.2$^{\circ}$; V$_{tot}$ 
$\simeq 26.5-0.025 X$\kms), where $X$ is distance in the direction
of the galactic center.

\subsection{Kinematics of the Sco-Cen OB association}

The Sco-Cen OB association (Sco OB2) and its three primary subgroups
(Upper Sco = US = Sco OB2\_2, Upper Centaurus Lupus = UCL = Sco OB2\_3, 
Lower Centaurus Crux = LCC = Sco OB2\_4) compromise the closest, large-scale 
site of recent high-mass star-formation to the Sun ($d \simeq 118-145$ pc; 
Z99). Kinematics of the three primary 
subgroups of the Sco-Cen association are given by the detailed studies 
of Z99, \cite{deBruijne00}, and \cite{Hoogerwerf00}. 
Z99 calculated the subgroup Galactic motions for 180 LCC, 221 UCL, 
and 120 US members using the median of the candidate's radial velocities 
listed in the {\it Hipparcos\,} Input Catalog \cite{tur93}.

Here, we calculate centroid positions and 
Galactic motion vectors to compare trajectories between these
massive subgroups and other smaller groups of young stars in
our discussion. Following Z99, we use mean proper motions and parallaxes,
and the median subgroup radial velocities. 
We used the {\it Hipparcos} mean right ascensions, 
declinations, and proper motions in those two directions, for all
the stars in Z99's membership lists. We adopt
their subgroup distances as defined by the member candidates of 
all spectral types, and the median radial velocity for all of the 
candidate members with radial velocities in the {\it Hipparcos} Input Catalog.
The Galactic motion vectors we calculated are purely heliocentric, 
and differ slightly from
those of Z99 (their Table A1).

\subsection{Kinematics of stars near $\epsilon$ Cha}

A kinematic group of X-ray emitting young stars at $d \simeq 110$ pc
in the vicinity of $\epsilon$ Cha was recently discovered in Chamaeleon by 
Frink et al.\ (1998, their
``subgroup 2'') and later discussed by Terranegra et al.\ (1999,
their ``kinematic group'').  \cite{egg98} independently pointed out a
grouping of early-type Local Association members between the Cha I, Cha
II, and Musca dark clouds, at similar distances.

We combined the lists of related stars from all 3 papers and found 
6 which had both {\it Hipparcos\,} astrometry and measured radial 
velocities, called here the ``$\epsilon$ Cha group''.  $\epsilon$ 
Cha AB (= HIP 58484) is a $V$ = 4.9 visual binary with a B9Vn primary.
Furthermore, $\epsilon$ Cha is comoving with the well-studied
\prems\, Herbig Ae/Be star HD 104237 (age $t \simeq 5$ Myr;
\cite{Terranegra99}) located 2.2$^{\prime}$ away.  $\epsilon$ Cha has
$v_r$\,=\,+13 $\pm$ 5 \kms (\cite{BB2000}); however HD 104237 has no 
published radial velocity and was ignored in our analysis. Due to 
the binarity of $\epsilon$ Cha, we adopt the Tycho-2 proper motion, 
but keep the {\it Hipparcos} parallax. \cite{Knee96} detected molecular gas
at $v_r$\,=\,+3.5\,$\pm$\,1.4 \kms in the close vicinity of $\epsilon$ Cha 
and HD 104237 and argue that the clouds are associated with the two 
young early-type stars due to the high ratio of IRAS 100\,$\mu$m dust 
emission to CO($J=1\,\rightarrow$\,0) emission. The velocity of the gas 
is closely bracketed by the radial velocity of gas in the Cha I and Cha II 
clouds, 
which are $d$ = 150-200 pc distant (\cite{Knude98} and references therein). 
Hence the cloudlets near $\epsilon$ Cha ($d = 112\,\pm\,7$\,pc; 
{\it Hipparcos}) appear to share the radial velocity of the main Cha 
clouds $50-100$ pc beyond it, and are 2$\sigma$ below the poor
radial velocity ($\pm$\,5\,\kms) for $\epsilon$ Cha in the literature. 
A more accurate radial velocity for $\epsilon$ Cha and HD 104237 
is desired to clear up this situation. However, for our purposes 
we use the Barbier-Brossat \& Figon radial velocity value.

Two RASS WTTs discovered by \cite{Alcala95} have {\it Hipparcos} 
entries and radial velocities 
from \cite{Covino97}, DW Cha (=RX J1158.5-7754a, HIP 54800; 
$v_r\,=\,+13.1 \pm 2.0$
\kms) and RX J1159.7-7601 (=HIP 58490; $v_r = +13.1 \pm 2.0$ \kms),
with estimated ages of $t \simeq 6$ Myr and 15 Myr, respectively
(\cite{Terranegra99}). DW Cha is binary, so we use the Tycho-2 proper
motion.

The isolated G8 ``weak-line YY Orionis star'' T~Cha (\cite{Alcala93}) 
had two
published radial velocities: $v_r\,=\,+14.6 \pm 2.1$ \kms (\cite{fra92}) and
$+20 \pm 2$ \kms (\cite{Covino97}), from which we adopt the weighted
mean $+17.4 \pm 1.5$ \kms. Terranegra et al.\ (1999) find an age $t
\simeq 30$ Myr for this G8 star. Its proper motion is similar
to the previously mentioned group members, however {\it Hipparcos} found
it to be much closer ($66^{+19}_{-12}$ pc). The galactic motion 
vector we compute for T Cha, (U,V,W) = $-2.3, -18.8, -10.5$ \kms, 
($\sigma_U$, $\sigma_V$, $\sigma_W$)
= 2.4, 1.6, 1.7 \kms, differs by $\approx$10 \kms in the X direction
from the other group members. Due to its age and discordent kinematics,
we exclude it as a member and discuss it no further. 

We added HD 104036 as a candidate group member (=HIP 58410, A7V, $V$
= 6.7), a star 24$^{\prime}$ NW of $\epsilon$ Cha and apparently
codistant and comoving with it. \cite{gre99} report a single observation
of $v_r$ = $+12.4\,\pm\,1.0$\,\kms for this star, consistent with the 
others.  On the Hertzsprung-Russell diagram, HD 104036 is above the
MS, close to the RS Cha binary and HD 104237, and appears to 
be a $t \simeq 6$ Myr star with $M \approx 2$ \msol 
(using D'Antona \& Mazzitelli 1997 tracks).

The Galactic locations and motions for the four likely members with
{\it Hipparcos} data and radial velocities are included 
in Table \ref{Table_KINE}. For the average position and motion of the 
group, we use the position of $\eta$ Cha as the center and the weighted 
mean UVW velocities of the four likely members with {\it Hipparcos} and 
$v_r$s ($\epsilon$ Cha, DW Cha, RX J1159.7-7601, HD 104036). The mean 
galactic motion vector implies that the group's convergent point should 
be near ($\ell$,$b$) = 301.3$^{\circ}$, --25.8$^{\circ}$, with a mean 
group velocity of 23.8\,\kms. This is similar to Eggen's Local 
Association or Pleiades Supercluster \cite{Eggen95} and the 
Gould Belt (\cite{Makarov00}) (\S 5.1).

During our analysis, we searched a 3$^{\circ}$ radius circle
around $\epsilon$ Cha in the Tycho-2 catalog for additional comoving
candidates. The four known members are spatially within 0.5$^{\circ}$ of each 
other, and form a tight proper motion locus.
Three additional candidates were identified in the Tycho-2 catalog: 
the A7III/IV star HD 105234
(=HIP 59093, $d = 106\,\pm\,7$ pc), the F0 star TYC 9420-676-1 (PPM 785582), and
TYC 9414-191-1. HR diagram placement of HD 105234 places it slightly above
the main sequence, bringing into question its luminosity classification, but
supporting the possibility of it being \prems. Spectroscopy of these candidates
is required to test for youth to see if they are indeed related to 
the $\epsilon$ Cha group.

\subsection{Kinematics of the TW Hydrae Association}

A fascinating group of nearby \prems\, stars in the fourth Galactic quadrant is
the TW Hydrae association (\cite{Webb99} and references therein).
\cite{sod98} showed preliminary evidence that one of its members, 
HD 98800, may be kinematically linked to the Sco-Cen OB association.  
We incorporate four of the TW Hya group's stars into our kinematic study 
-- TW~Hya, HD 98800, HR 4796, and TWA 9 -- which have accurate {\it Hipparcos} 
proper motions and parallaxes, along with published radial velocities.

For TW Hya we adopt the very precise radial velocity presented by 
\cite{Torres99} ($v_r = +12.9\,\pm\,0.2$ \kms). 
For HD 98800, the mean of the $\gamma$ values from \cite{Torres95} for the 
Aa+Ab and Ba+Bb pairs were used, and assigned a 0.2 \kms\, 
uncertainty (where $\gamma$ is the center-of-mass radial velocity
of a stellar pair in an orbit solution). 
Due to the multiplicity of HD 98800,
we use the long-baseline proper motion from the Tycho-2 catalog over 
the {\it Hipparcos} proper motion, but of course use the {\it Hipparcos}
parallax. 

For HR 4796A, we used the value
given by \cite{BB2000} ($v_r = +9.4\,\pm\,2.3$ \kms) which represents
the average of 7 observations. \cite{Stauffer95} found $v_r = +9 \pm 1$ \kms 
for the M2.5 companion HR 4796B, agreeing closely with the HR 4796A value. 
We use the Tycho-2 proper motion since HR 4796 is binary, 
although it agrees well with the {\it Hipparcos} value.

The radial velocity for TWA 9 was reported by \cite{Torres99}: 
$v_r = +9.9\,\pm\,0.4$ \kms. Both TWA 9 A and B are suspiciously
underluminous for Li-rich stars (see Fig. 3 of Webb et al. 1999),
and their proper motions are smaller than most of the other TWA members.
We propose that the {\it Hipparcos} parallax may be underestimating
the distance to TWA 9 due to multiplicity, and $d = 70-80$ pc may be 
more appropriate,
however we retain the {\it Hipparcos} parallax in our calculations.
The factor of 2.2 in the spread of the TW Hya stellar proper motions
in Webb et al.'s study suggest that the association has a factor
of $\sim$2.2 depth, i.e. the members likely have distances of 
$\sim 30-70$ pc.
This is likely to be a major source of scatter in Webb et al.'s 
HR diagram, comparable in magnitude to the effects of stellar multiplicity.

The weighted mean locations and velocities of these three stars are
given in Table \ref{Table_KINE}. We treat TW Hya itself as the
center of the association (this is appropriate considering
the distribution of members in Webb et al.'s Fig. 2, and the mean
distance of $\sim$50 pc). The association's
galactic motion vector is a standard mean of the values for the 4 members 
considered here, and is consistent with a convergent point of 
($\ell$,$b$) = 241$^{\circ}$, --12$^{\circ}$ and total
motion of 21 \kms. As with the $\eta$ Cha and $\epsilon$ Cha groups,
this motion resembles that of Eggen's Local Association or 
Pleiades Supercluster and the Gould Belt population.

\section{Notes on confirmed $\eta$ Cha members from Paper I \label{appendix}}

\subsection{RECX 1 = RX J0837.0--7856 = GSC 9402\_00921 = DENIS
J083656.5-785646} 

High-resolution spectroscopy of RECX 1 indicates a K4 spectral type 
with a Li $\lambda$6707 equivalent width $EW$(Li) = 0.52\AA~and 
rotational velocity \vsini= 13 $\pm$ 3 \kms (\cite{Covino97}).  We 
use the $V$ magnitude from Tycho ($V$ = 10.52;
\cite{ESA97}) and the bolometric correction (BC) for a K4 dwarf from
Kenyon \& Hartmann (1995), assuming zero reddening, to calculate the
bolometric luminosity log($L$/\lsol) = $-$0.12.  Its $v_r = +18\pm2$ \kms
(Covino et al. 1997) is similar to that for $\eta$ Cha and RS Cha, and its
space motion is parallel to $\eta$ Cha, RS Cha, HD 75505, and RECX 11.
This star had the highest X-ray luminosity of the RECX stars, and
appears double ($\approx 8\arcsec.6$, 6$^{\circ}$ NE) but saturated in
the digitized Sky Survey images.  The DENIS sampler (\cite{Epchtein99})
lists preliminary photometry of $Ks_{\rm auto}$ = 7.22.

\cite{fri98} include the star in their proper motion subgroup \#2 whose
other members are all Cha I and Cha II T Tauri stars.  If RECX 1 had
formed in Cha I at $d \simeq 140$ pc,  it would be $\simeq 1$ Myr old
and would have traversed at least 19 pc (7.8$^{\circ}$ projected),
requiring a speed of at least 18\,\kms with respect to its parent
cloud. Similarly, to originate in Cha II ($d$ = 200 pc), RECX 1 would be
$\sim$0.5 Myr old and would need a relative speed of 53 \kms.  Since
RECX 1 is clearly a $\eta$ Cha cluster member, birth in Cha I or Cha II
seems implausible.

\subsection{RECX 2 = $\eta$ Cha = HD 75416 = HR 3502 = \iras
F0843-7846}

$\eta$ Cha is the brightest ($V$ = 5.46; Tycho) and earliest type (B8V;
\cite{hou75}) of the co-moving intermediate-mass stars in the cluster.
Historically, the 4th edition of the Bright Star Catalogue lists 
$\eta$ Cha as a member of
the Sco-Cen Association (\cite{hof82}), however it is well
outside of the classically defined association region.
Its {\it Hipparcos} \BV\,  color index matches that expected for a B8
star within 0.01 mags, so reddening is negligible in agreement with
Stromgren photometric properties E($b-y$) = $-0.004$, log $T_{\rm eff}$ =
4.107, dM$_{b}$=0.183 (\cite{Westin85}). Using the {\it Hipparcos}
distance and $V$ magnitude and no reddening, we calculate a luminosity
of log($L$/\lsol) = 2.02, placing it directly on the solar metallicity main
sequence as given in both \cite{Pols97} and \cite{sch92}.  A 3.8
\msol\, star reaches the main sequence with spectral type B8V in
$1-2$ Myr (\cite{pal93}). 

Mannings \& Barlow (1998) found $\eta$ Cha to be a candidate main
sequence star with a debris disk or a ``Vega-like source''. This is
reminiscent of HR 4796A, the well-studied $\sim$10-Myr-old member of
the TW Hya association with an infrared excess due to circumstellar
material. We found the star to be an X-ray emitter 
(log \lx ~= 28.7~\ergs), unusual
for late B stars (e.g.  \cite{gri92}).   Spectroscopic studies of
$\eta$ Cha found it to be a radial velocity variable ranging from $-17$
\kms to $+37$ \kms (\cite{bus61}, \cite{neu30}), thus it is likely a
binary.  The X-ray emission is likely to be produced by the lower-mass
secondary rather than the intermediate-mass primary (\cite{Simon95}).

\subsection{RECX 3 = DENIS J084137.2--790331}

The DENIS sampler (\cite{Epchtein99}) lists preliminary photometry of
Gunn $i_{\rm auto}$} = 11.57, $Ks_{\rm auto}$ = 9.47.

\subsection{RECX 4 = GSC 9403\_1083 = DENIS J084224.0--790403}

The DENIS sampler (\cite{Epchtein99}) lists preliminary photometry of
$J_{\rm auto}$ = 9.54, $Ks_{\rm auto}$ = 8.66.

\subsection{RECX 5 = DENIS J084120.1--785751}

The DENIS sampler (\cite{Epchtein99}) lists preliminary photometry of
Gunn $i_{\rm auto}$ = 6.57, $J_{\rm auto}$ = 7.44.

\subsection{RECX 7 = RX J0842.9--7904}

This star has neither a GSC or USNO-A2.0 entry, apparently due to its
proximity (40\arcsec~SW) to the bright binary RS Cha.  Its position
in Paper I was estimated from DSS, and its $F$ magnitude was estimated
from comparisons with other nearby USNO stars.  On POSS-II, it appears
as a visual binary (separation 11$^{\arcsec}$.4, {\it PA\,} $\simeq
30^{\circ}$).  RECX 7 may be also be related to RECX 8 (RS Cha).  The 
star has a double H$\alpha$ profile (Paper I) which may indicate that 
the star is a spectroscopic binary, or that combination of emission 
and absorption processes contribute to the line profile (\cite{rei96}).
High-resolution spectroscopy shows a K4 spectral type,
$EW$(Li) = 0.46\AA, rotational velocity \vsini\, = $28-32$ \kms, and $v_r =
+4.3\pm2$ \kms (\cite{Covino97}).  A spectral type of  K3 and $EW$(Li)
= 0.4 \AA\, were derived from a lower resolution spectrum in Paper I.

\subsection{RECX 8 = RS Cha AB = HD 75747 = HR 3524}

RS Cha AB is a A8V double-lined eclipsing binary that has been
well-studied since its discovery as a variable star by \cite{str64}.
Accurate physical parameters of the binary are from a spectroscopic
study by Andersen (1975) and detailed 4-color $uvby$ photometry 
by Clausen \& Nordstr\"om (1980, hereafter CN80).  
Andersen also suggested the $\delta$ Scuti 
nature of the primary.  In the comprehensive review of detached,
double-lined eclipsing binary systems, Andersen (1991, hereafter
A91) listed
revised component masses of 1.858\,$\pm$\,0.016 
and 1.821 $\pm$ 0.018 \msol, for A and B respectively, 
and temperatures $T_{\rm eff}$(A) = 8050 $\pm$ 200\,K and
$T_{\rm eff}$(B) = 7700 $\pm$ 200\,K, and radii from 
CN80.  A91 calculated identical luminosities 
of log($L$/\lsol) = 1.24\,$\pm$\,0.03 for A and B, independent 
of distance or bolometric correction.

In the post-{\it Hipparcos} era, two studies have reexamined the
system as part of a sample of eclipsing binaries to test current 
calibrations of temperatures (Ribas et al. 1998; hereafter R98) 
and photometric estimates 
of surface gravities (Jordi et al. 1997; hereafter J97). R98 used the 
accurate {\it Hipparcos} parallax and component radii to calculate 
effective temperatures of $T_{\rm eff}$(A) = 7687\,$\pm$\,180 K and 
$T_{\rm eff}$(B) = 7331\,$\pm$\,170 K.  (Note that the R98
 estimates are dependent on the bolometric corrections 
of \cite{Flower96}, which are +0.03 for these late A stars.)  
Using the CN80 photometry, J97 found
$T_{\rm eff}$(A) = 7810\,K and $T_{\rm eff}$(B) = 7295\,K 
by interpolating the Str\"omgren-Crawford photometry and
atmosphere model parameter grids of \cite{Moon85}.  Although
they do not give errors, we conservatively assume a 200 K 
uncertainty similar to those of R98 and CN80.  
After reviewing all of the available literature, 
we adopt the new J97 temperatures and the CN80 radii, 
and calculate new luminosities of
log($L$/\lsol) = 1.182\,$\pm$\,0.047 for A, and 
log($L$/\lsol) = 1.142\,$\pm$\,0.049 for B 
(assuming T$_{\odot}$ = 5781 K as adopted by \cite{Bessell98}).

All previous estimates for the age of this binary come from 
{\it post}-main sequence stellar evolution models inferring an 
age of $8-10 \times 10^{8}$ yrs.  \cite{Pols97} found that 
``{\it it is not possible to fit the masses and radii of both stars 
[RS Cha AB] at the same age, for any metallicity,}'' and 
that an implausible 
age difference of $\sim$100 Myr is required to reconcile the 
high luminosity of the secondary.

Solving the discordance in the post-MS ages, Paper I found 
that the system is \prems.  We plot the system on an 
Hertzsprung-Russell diagram with the pre-MS evolutionary 
tracks of D'Antona \& Mazzitelli (private communication, 
hereafter DM00) in 
Figure \ref{Fig_RSCHA}, comparing the temperatures and 
luminosities given by A91 and our values (labeled ``New'')
using the J97 temperatures and CN80 radii. 
DM00 kindly provided evolutionary tracks for masses
$1.7-2.0$ \msol~ in 0.1 \msol steps, and between $5-9$
Myr in 1 Myr steps. We also inferred ages and masses
from the isochrones of Palla \& Stahler 
(1999; hereafter PS99). Our analysis is not meant to 
distinguish between the full spectrum of evolutionary 
models currently available, of which we have
included only two, but to illustrate that 
the components of RS Cha are consistent with 
being coeval \prems\, stars.

We derive model masses from the
evolutionary tracks, and give approximate 1$\sigma$ 
uncertainties. The upper(lower) 1$\sigma$ values of the
model ages were interpolated using the edge of the
(T$_{eff}$, log(L)) error oval closest to the 
next oldest(youngest) isochrone. 
  
By interpolation of the DM00 tracks in 
Fig. \ref{Fig_RSCHA} we derive model masses and ages of
1.82\,$\pm$\,0.04 \msol~and 7.5\,$\pm$\,0.5 Myr for A, and
1.79\,$\pm$\,0.05 \msol~and 7.7\,$\pm$\,0.7 Myr for B, using
our new positions. 
With the A91 temperatures and luminosities and the DM00 tracks,
we find 
1.88\,$\pm$\,0.04 \msol~and 7.2\,$\pm$\,0.5 Myr for A, and
1.88\,$\pm$\,0.05 \msol~and 7.0\,$\pm$\,0.5 Myr for B.
On the pre-MS tracks of PS99,we
interpolated between their 1.5\,\msol\, and 2.0\,\msol\,
tracks to extract masses and ages.
For the A91 position on the PS99 tracks, we find 
1.85\,$\pm$\,0.03 \msol~and 4.7\,$\pm$\,1.0 Myr for A, and
1.88\,$\pm$\,0.02 \msol~and 3.5\,$\pm$\,0.7 Myr for B.
For our new (T$_{eff}$, log(L)) values on the PS tracks we find 
1.82\,$\pm$\,0.04 \msol~and 4.5\,$\pm$\,1.4 Myr for A, and
1.81\,$\pm$\,0.04 \msol~and 4.0\,$\pm$\,0.9 Myr for B.
The differences in the ages for the binary were
nearly a factor of 2 ($\approx$7 Myr for DM00 and
$\approx$4 Myr for PS99), however the masses
were remarkably close to the dynamical masses of A91.

With our RS Cha HR diagram positions, the DM00 pre-MS model 
masses differ by --2\% from Andersen's dynamical 
masses for both components, while the model masses with the A91 
temperatures and luminosities are +1 and +3\% higher, 
for A and B respectively. 
The PS99 model masses with the new HR diagram
positions are 0\% and 3\% higher than the dynamical masses,
and 1\% and 2\% lower for the A91 temperature and luminosity,
for A and B respectively. Within our 1$\sigma$ model age
uncertainties, the RS Cha binary is consistent with being
coeval for both the DM00 and PS99 evolutionary models, 
thereby resolving the 
age discrepancy reported by \cite{Pols97}. 
RS Cha should be an important benchmark binary for testing 
evolutionary models of \prems\, intermediate mass stars.

\placefigure{Fig_RSCHA}

As a tight eclipsing binary with period 1.7 days, the RS Cha system is
estimated to circularize its orbit in about 5 Myr (\cite{may91}). The
orbit is observed to be circular, consistent with our age estimate of
the $\eta$ Cha cluster.  Detailed modeling of the individual stars and
the system dynamics may be useful in constraining pre-MS evolutionary
models.

Our HRI observation of RS Cha provides one of the longest X-ray light
curves of a pre-MS A star ever obtained.  X-ray emission seen in some
A stars, which have quiescent atmospheres that ought not generate
solar-type magnetic activity, has been a mystery for nearly two decades
(\cite{Simon95} and references therein).  The RS Cha X-ray flare-like
variations are quite similar to those seen in the late-type $\eta$ Cha
sources (Figure \ref{Fig_XRAY})\footnote{RS Cha was reported at a
higher level of $3 \times 10^{30}$ erg s$^{-1}$ during the RASS survey
(\cite{Huensch98}) but this may represent confusion between RECX 8 (=
RS Cha) and the X-ray brighter star RECX 7 (= RX J0842.9--7904) which 
is less than 1\arcmin~away.}.  These characteristics suggest that 
RS Cha AB is actually a triple system where the X-ray emission is due 
to an undetected low-mass companion.

\subsection{RECX 9}

Paper I noted the star had a double H$\alpha$ profile.  As for RECX 7,
this could indicate a double-lined spectroscopic binary or due to a 
mixture of emission and intervening absorption components (\cite{rei96}).

\subsection{RECX 10 = RX J0844.5--7846 = GSC 9403\_1279}

One of the WTTs discovered with the ROSAT All-Sky Survey, this star has
spectral type K7--M0, $EW$(Li) =  0.52 \AA, \vsini = $9 \pm 5$ \kms
and $v_r = +15.0$ \kms (Paper I, Covino et al. 1997).

\subsection{RECX 11 = NSV 4280 = BV 1051 = \iras\/ F08487--7848 =
	    1RXS J084659.3--785938 = GSC 9403\_1016}

RECX 11 is listed in the \iras\/ Faint Source Catalogue (\cite{mos89})
and suspected variable star catalog (\cite{kaz98}), and was detected 
in X-rays by the {\it ROSAT\,} All-Sky Survey (Alcal\'a et al. 1995) 
and our pointed HRI observation. The DENIS sampler (\cite{Epchtein99}) gives
preliminary photometry of $i_{\rm auto}$ = 9.75, $J_{\rm auto}$ = 8.76,
$Ks_{\rm auto}$ = 7.65 which, assuming K7 spectral type, indicates $\Delta
(J$--$Ks) \simeq 0.3$ color excess.  It thus appears to be a
transitional Class II/III young stellar object between the classical to
weak-lined T Tauri phases with a heated dusty disk but only weak
H$\alpha$ emission suggesting that active accretion has ceased.  A
well-studied star at a similar transitional phase, though with a
younger age around 1 Myr, is HD 283447 in the Taurus cloud complex
(e.g. \cite{Feigelson94}).

\subsection{RECX 12 = RX J0848.0--7854 = GSC 9403\_0489 = DENIS
J084757.1--785454}

One of the 4 previously known RASS WTTs in the region, this is a M2 star
with $EW$(Li) = 0.61 \AA, \vsini = $13 \pm 3$ \kms and $v_r = +18.0$
\kms (Paper I, Covino et al. 1997).  The DENIS sampler (\cite{Epchtein99}) 
gives preliminary photometry of Gunn $i_{\rm auto}$ = 10.63, 
$J_{\rm auto}$ = 9.26, $Ks_{\rm auto}$ = 8.36.

\subsection{HD 75505 = GSC 9403\_0119}

HD 75505 (Tycho $V_T$ = 7.41, Johnson $V = 7.39$) was the only 
bright AB star in the 
cluster core not detected by \rosat\/ HRI. Tycho gives a 
low-quality parallax $\pi = 11.9 \pm 2.6$ mas or 
$d = 84^{+24}_{-15}$ pc, consistent
with that of $\eta$ Cha and RS Cha. The star is classified 
A1V (\cite{hou75}) but its Tycho $(B-V)_T$ color infers
a Johnson color index of \bv = 0.158, about 0.13 mags 
redder than an unreddened A1 star (\cite{ken95}) but appropriate for 
an unreddened A6 dwarf.  If the spectral type and Tycho color 
is correct, then HD 75505 (with A$_{V}$ = 0.41) is much more 
reddened than $\eta$ Cha and RS Cha (A$_{V}$ $\simeq$ 0.00), 
suggesting unrealistic clumpy extinction within
the cluster core.  Adopting the cluster distance $d = 97$ pc and
correcting for $A_V=0.41$ extinction, HD 75505 falls 0.05 dex in
luminosity below the ZAMS (i.e. 100 Myr isochrone) for a $\sim$2
\msol\, star, which is much older than the other stars in the
$\eta$ Cha cluster.

To resolve this issue, we examined the preliminary 
photometry from the DENIS sampler (\cite{Epchtein99}),
which lists Gunn\,$i_{\rm auto}$ = 7.89, 
$J_{\rm auto}$ = 7.07, $Ks_{\rm auto}$ = 6.93 for HD 75505.  
Assuming the DENIS $J$ and $Ks$ magnitudes are similar to Johnson
$J$ and $K$, the $(V - J)$ and $(V - Ks)$ colors 
are both consistent with unreddened A6 dwarf colors. Hence
the optical and near-IR colors are both consistent with
HD 75505 being an A6 dwarf. With an A6 spectral type,
a distance of 97\,pc, and no extinction, HD 75505 lies within
0.1 dex of log(L) of the ZAMS in DM97's models. Its position
on the ZAMS implies a {\it minimum} age of about $8-10$ Myr, consistent
with the model ages of other members in Paper I.

\appendix

\clearpage

\begin{figure}
\begin{center}
\epsfxsize=8cm
\epsffile{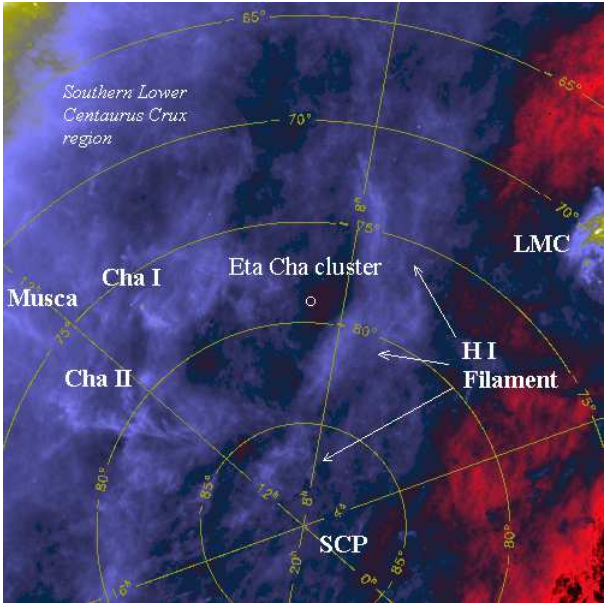}
\caption{
\iras ~ 100\,$\mu$m image, $30^{\circ}$ in extent, 
of the $\eta$ Cha cluster vicinity. The dust feature associated with
the H {\small I} filament, important
molecular clouds in the region, and the Sco-Cen LCC current location
are indicated.  
\label{Fig_IRAS}
}
\end{center}
\end{figure}


\begin{figure}
\begin{center}
\epsfxsize=8cm
\epsffile{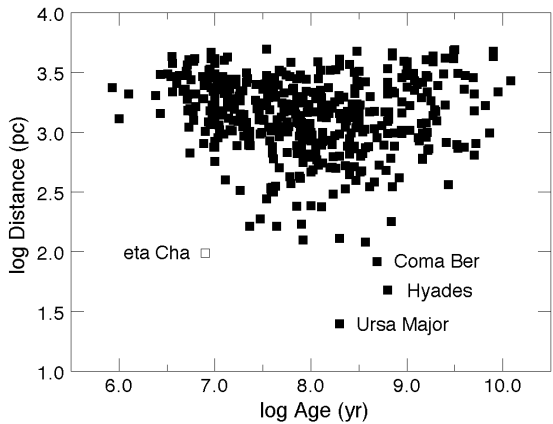}
\caption{
Comparison between the $\eta$ Cha cluster and 480 clusters
with $d <$ 5000 pc (\cite{Mermilliod95}) showing it is unusually
close and young. 
\label{Fig_LYNGA}
}
\end{center}
\end{figure}


\begin{figure}
\begin{center}
\epsfxsize=8cm
\epsffile{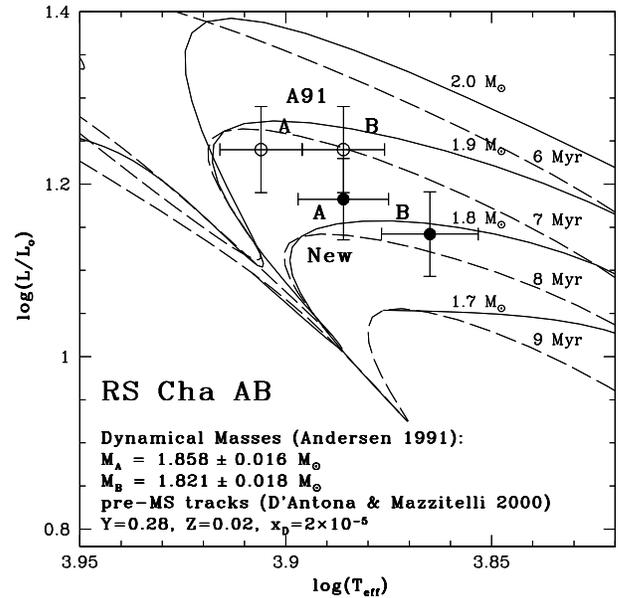}
\caption{
Theoretical HR diagram position for the components of the 
RS Cha eclipsing binary system. The isochrones of DM00
are overlayed. The difference between the Andersen (1991)
positions and the new positions we calculated are that 
we used the temperatures of Jordi et al. (1997; J97) 
(see Appendix B.7). The dynamical masses and pre-MS model
masses agree within 3\% for the tracks of DM00 and PS99.
The system appears to be a coeval pre-MS binary with age 
estimates of 7--8 Myr (DM00) or 4--5 Myr (PS99). 
\label{Fig_RSCHA}
}
\end{center}
\end{figure}

\clearpage

\begin{deluxetable}{crrccccrr}
\footnotesize
\tablenum{1}
\tablewidth{420pt}
\tablecaption{\rosat\/ $\eta$ Cha X-Ray sources \label{Table_XRAY}}
\tablehead{
       & \multicolumn{2}{c}{Count rate} & \colhead{$\log(L_X)$\tablenotemark{a}} & 
\colhead{Stellar} & \colhead{Offset} & \colhead{Variability} & 
\multicolumn{2}{c}{USNO photometry} \\ 
\cline{2-3}
\cline{8-9}
RECX   & \colhead{cts ks$^{-1}$} & \colhead{$\pm$} & \colhead{erg s$^{-1}$} & 
\colhead{counterpart} & \colhead{\arcsec} & \colhead{cts ks$^{-1}$} &
\colhead{\bj} & \colhead{$F$} \\
\colhead{(1)} & \colhead{(2)} & \colhead{(3)} & \colhead{(4)} & \colhead{(5)}
& \colhead{(6)} & \colhead{(7)} & \colhead{(8)} & \colhead{(9)}
 }
\startdata
 1 & 105~~~~~~ & 2~~ & 30.6 & GSC 9402\_0921 & 2.1 & Yes (0.08--0.22) & 11.5 & ~9.7 \\
 2 &   1.2~~~~ & 0.3 & 28.8 & $\eta$ Cha      & 0.9 & No  & 
       ~...\tablenotemark{c} & ~...\tablenotemark{c}\\
 3 &   3.2~~~~ & 0.4 & 29.2 & USNO-A2.0       & 1.2 & No               & 12.6 & 14.0 \\
 4 &  32~~~~~~ & 1~~ & 30.2 & GSC 9403\_1083 & 1.5 & Yes (0.01--0.04) & 14.4 & 11.6 \\
 5 &   2.5~~~~ & 0.3 & 29.1 & USNO-A2.0       & 1.6 & No               & 17.6 & 13.9 \\
 6 &   8.2~~~~ & 0.5 & 29.6 & GSC 9403\_0288 & 3.6 & Yes (0.01--0.04) & 15.6 & 12.6 \\
 7 &  50~~~~~~ & 1~~ & 30.4 & anonymous       & 2.7 & Yes (0.04--0.09) & 
       ~...\tablenotemark{d} & 10:\tablenotemark{d}\\
 8 &  17.0~~~~ & 0.7 & 29.9 & RS Cha          & 2.8 & Yes (0.01--0.05) &
       ~...\tablenotemark{c} &  ...\tablenotemark{c}\\
 9 &   0.6~~~~ & 0.3 & 28.5 & USNO-A2.0       & 1.0 & No               & 15.2 & 13.4 \\
10 &  17.9~~~~ & 0.8 & 30.0 & GSC 9403\_1279 & 3.7 & Possible         & 13.5 & 11.4 \\ 
11 &  32~~~~~~ & 1~~ & 30.2 & GSC 9403\_1016 & 3.3 & Yes (0.02--0.14) & 12.1 & 10.8 \\ 
12 &  33~~~~~~ & 1~~ & 30.2 & GSC 9403\_0389 & 4.0 & Yes (0.02--0.11) & 13.9 & 11.8 \\
... &   $b$~~~~ & $b$ & $b$  & GSC 9398\_0105 & $b$ & $b$              & 12.3 & 11.6 \\ 
\enddata
\tablenotetext{a}{Log ($L_x$) values are 0.1 above those reported in Paper I
due to an earlier error in determining the exposure time corrections.}
\tablenotetext{b}{Faint source close to northern edge of HRI field where 
exposure time is uncertain due to satellite dithering.  The optical
counterpart is too faint and blue to be a cluster member, hence we do 
not assign it an RECX number.}   
\tablenotetext{c}{Unreliable magnitudes from Sky Survey Schmidt plates 
due to saturation.}
\tablenotetext{d}{Star is not listed in catalogs due to proximity to 
RS Cha.  $F$ magnitude is estimated by visual inspection of DSS.} 
\end{deluxetable}

\clearpage

\begin{deluxetable}{ll}
\footnotesize
\tablenum{2}
\tablewidth{240pt}
\tablecaption{$\eta$ Cha cluster properties \label{Table_LYNGA}}
\tablehead{\colhead{Property} & \colhead{Value} \\
           \colhead{(1)} & \colhead{(2)}}
\startdata
Designation               & C J0842--790 \\
Location (J2000)          & 8$^{\rm h}$ 42$^{\rm m}$ 06$^{\rm s}$, 
                            --79$^{\circ}$ 01$\arcmin$ 38$\arcsec$ \\
~~~~~~~~ (Galactic)       & 292.48$^{\circ}$, --21.65$^{\circ}$ \\
Parallax (distance)       & 10.28 $\pm$ 0.31 mas (97.3 $\pm$ 3.0 pc) \\
Brightest star            & $\eta$ Cha, $V = 5.46$, B8V \\
$E$\BV                    & 0.00 \\
Angular (linear) diameter & 40\arcmin: (1.2: pc) \\
log(Age) (yr)             & 6.9 $\pm$ 0.3 \\
Membership                & 50? (known primaries: 13)\\
Mass                      & $\sim 23$ \msol (known members: 13 \msol)\\
Trumpler class            & II3p  \\
Proper motion ($\alpha$)  & $\mu_{\alpha}$ = --30.0\,$\pm$\,0.3 mas yr$^{-1}$\\ 
Proper motion ($\delta$)  & $\mu_{\delta}$ = --27.8\,$\pm$\,0.3 mas yr$^{-1}$\\
Radial velocity           & +16.1 $\pm$ 0.5 km s$^{-1}$ \\
Heliocentric Position (X,Y,Z) & (+35, --84, --36) pc \\
Heliocentric Velocity (U,V,W) & (--11.8, --19.1, --10.5) km s$^{-1}$ \\
\enddata
\end{deluxetable}

\clearpage

\begin{deluxetable}{ccccccccccc}
\tiny
\tablenum{3}
\tablewidth{480pt}
\tablecaption{Galactic Positions and Kinematics of young stars and
stellar systems around Chamaeleon \label{Table_KINE}}
\tablehead{
\colhead{} & \colhead{X,$\sigma_X$} & \colhead{Y,$\sigma_Y$} & 
\colhead{Z,$\sigma_Z$} & \colhead{U,$\sigma_U$} & \colhead{V,$\sigma_V$} & 
\colhead{W,$\sigma_W$} & \colhead{Age\tablenotemark{a}} & \colhead{Closest} & 
\colhead{Time} & \colhead{$\Delta$} \\
\colhead{Object} & \colhead{(pc)} & \colhead{(pc)} & \colhead{(pc)} & 
\colhead{\kms} & \colhead{\kms} & \colhead{\kms} & \colhead{(Myr)} & 
\colhead{Group} & \colhead{(Myr)} & \colhead{(pc)} \\
\colhead{(1)} & \colhead{(2)} & \colhead{(3)} & \colhead{(4)} & 
\colhead{(5)} & \colhead{(6)} & \colhead{(7)} & \colhead{(8)} & 
\colhead{(9)} & \colhead{(10)} & \colhead{(11)} 
}
\startdata
$\eta$ Cha Cluster Members\\
\\
$\eta$ Cha &  34.3(1.4) &  --83.3(3.5) & --35.8(1.5) & --12.2(3.6) & 
--17.2(8.6) &  --9.6(3.7) & MS & LCC & --10 & 24 \\
RS Cha &  34.9(1.6) &  --83.9(3.8) & --36.0(1.6) & --12.3(0.8) & 
--19.1(0.5) & --10.6(0.4) & 8  & LCC & --10 & 17 \\
RECX 1\tablenotemark{b} &  34.2(1.0) &  --83.6(2.5) & --36.2(1.1) & 
--10.3(1.2) & --20.4(1.8) & --11.1(1.1) & 6  & LCC & --11 & 24 \\
\\
TW Hya Association Members\\
\\
TW Hya & 7.8(1.0) &  --51.4(6.4) &  22.0(2.7) & --12.0(1.8) & --18.2(0.9) &  
--5.0(1.4) & 10    & LCC & --14 & 23 \\
HD 98800 &  5.7(0.8) &  --38.4(5.1) &  26.0(3.5) & --13.3(1.9) & --17.9(1.4) & 
--6.9(1.6) & 10    & US & --16 & 18 \\
HR 4796  &  30.6(1.5) &  --53.7(2.7) &  26.1(1.3) & 
--8.5(1.3) & --18.3(1.9) &  --3.6(1.0) & 8 $\pm$ 2 & UCL & --18 & 28 \\
TWA 9 (CD\,--36$^{\circ}$\,7429)  &  15.2(1.8) &  --43.5(5.2) &  20.2(2.4) &
--7.0(1.3) & -15.2(0.9)  &  --3.0(1.0) & ?  & LCC & --6 & 63 \\
\\
$\epsilon$ Cha Group Candidate Members\\
\\
DW Cha &  41.5(4.6) &  --71.7(8.0) & --22.7(2.5) &  
--7.9(2.1) & --17.8(1.9) &  --7.9(1.2) & 6   & UCL & --14 & 38 \\
RXJ 1159.7-7601 	     &  44.6(7.1) &  --78.0(12.4)& --21.5(3.4) &  
--8.8(2.6) & --18.2(2.0) &  --8.5(1.2) & 15  & LCC & --16  & 14 \\
HD 104036 		     &  51.1(2.8) &  --88.3(4.9) & --27.8(1.5) & 
--12.2(1.1) & --18.1(0.9) & --11.9(0.6) & 6--7 & LCC & --7  & 17 \\
$\epsilon$ Cha (HD 104174)   &  54.1(3.5) &  --93.0(6.0) & --30.1(2.0) & 
--11.4(2.8) & --18.0(4.2) & --10.7(1.5) & 5? & UCL & --10  & 40 \\
\\
Clusters/Associations\\
\\
$\eta$ Cha Cluster\tablenotemark{c} & 34.6(1.1) &  --83.6(2.6) & --35.9(1.1) & 
--11.8(0.7) & --19.1(0.5) & --10.5(0.3) & 8\,$\pm$\,4 & LCC & --11 & 13 \\
TW Hya Assn.       & 54: & --93: & --30: & --10.2(2.9)  & --17.4(1.5) &  --4.6(1.8)&
$\approx$10 & LCC & --19 & 34 \\
$\epsilon$ Cha Group & 48: & --83: &  --26: & --10.2(2.2) & --18.6(1.1) &  --8.8(2.1)&
$\approx$5-15 & LCC & --13 & 22 \\

Lo. Cen. Crux (LCC) & 61.6(17.7)& --102.5(23.0)&  13.8(15.9)&  --8.8 & 
--20.0 &  --6.2 & 11--12 & UCL & --11 & 47 \\
Up. Cen. Lup. (UCL) & 121.8(29.8)&  --68.6(25.8)&  32.3(16.4)&  --3.9 & 
--20.3 &  --3.4 & 14--15 & US & --4 & 46 \\
Up. Sco. (US) & 141.2(34.3)&  --21.8(10.8)&  49.9(16.1)&  --4.7 & 
--16.8 &  --6.7 & 4--5   & UCL & --4 & 46 \\
\enddata
\tablenotetext{a}{
Ages of $\eta$ Cha stars from Paper I, Sco-Cen subgroups from 
de Geus et al. (1989), TW Hya and TWA 9 from Webb et al. (1999), HD 98000 
from \cite{sod98}, HR 4796 from Stauffer et al. (1995), T Cha, 
DW Cha, RX J1159.7-7601 from Terranegra et al. (1999), 
HD 104036 and $\epsilon$ Cha are from isochrone fitting done by 
us using \cite{DAntona97} pre-MS models. UVW galactic motion vectors
are calculated using {\it Hipparcos} and Tycho-2 astrometry and
radial velocities in the literature; see Appendix A for references.}
\tablenotetext{b}{
To calculate the position and velocity of RECX 1, we assumed $\pi$ to be
the weighted mean of the individual parallaxes for RS Cha and $\eta$ Cha.}
\tablenotetext{c}{
$\eta$ Cha cluster mean position is weighted mean for $\eta$ Cha and 
RS Cha; cluster mean velocity is calculated in Appendix A.1}
\end{deluxetable}

\clearpage
\end{document}